\documentclass[10pt,twocolumn]{IEEEtran}
\usepackage{graphicx}
\usepackage{subfigure}
\usepackage{color}
\usepackage{epsfig}
\usepackage{amsmath}
\usepackage{amssymb}
\usepackage{latexsym}
\usepackage{nicefrac}

\newcommand{\liminfty}[1]{\mbox{$\lim_{#1 \rightarrow \infty}$}}

\newcommand{\bE}[1]{{\mathbb{E}}\left[{#1}\right]}
\newcommand{\bP}[1]{{\mathbb{P}}\left[{#1}\right]}

\newcommand{\1}[1]{{\bf 1}\left[#1\right]}       % indicator 1{...}

\newcommand{\fsquare}{\vrule height6pt width7pt depth1pt}   % filled square
\newcommand{\myproof}{{\hfill \\ \bf Proof. \ }}           % Proof
\newcommand{\myendpf}{\hfill\fsquare \\[0.1in]}             % end of proof
\newcommand{\myvec}[1]{{\mbox{\boldmath{$#1$}}}}

\newcommand{\remove}[1]{}

\newtheorem{theorem}{Theorem}[section]

\newtheorem{lemma}[theorem]{Lemma}
\newtheorem{proposition}[theorem]{Proposition}

\renewcommand{\baselinestretch}{0.987}
\begin{document}

% paper title
\title{Intersecting random graphs and networks with
       multiple adjacency constraints: A simple example} %JSAC
% \title{Networks with multiple adjacency constraints: Beyond the disk model} %Infocom      

\author{N. Prasanth Anthapadmanabhan,~\IEEEmembership{Student Member, IEEE}
\thanks{N. Prasanth Anthapadmanabhan is with the Department 
       of Electrical and Computer Engineering, 
       and the Institute for Systems Research, University of Maryland, 
       College Park, MD 20742, U.S.A. E-mail: nagarajp@umd.edu.}
and
Armand M.  Makowski,~\IEEEmembership{Fellow, IEEE}
\thanks{Armand M. Makowski is with the Department of
Electrical and Computer Engineering, and the Institute for Systems
Research, University of Maryland, College Park, MD 20742, U.S.A.
E-mail: armand@isr.umd.edu.}
\thanks{This work was
prepared through collaborative participation in the Communications
and Networks Consortium sponsored by the U. S. Army Research
Laboratory under the Collaborative Technology Alliance Program,
Cooperative Agreement DAAD19-01-2-0011. The U. S. Government is
authorized to reproduce and distribute reprints for Government
purposes notwithstanding any copyright notation thereon.
%The views and conclusions contained in this document are those of
%the authors and should not be interpreted as representing the
%official policies, either expressed or implied, of the Army Research
%Laboratory or the U. S. Government.
} 
} 

\maketitle

\begin{abstract}
When studying networks using random graph models, 
one is sometimes faced with situations where
the notion of adjacency between nodes reflects multiple constraints. 
Traditional random graph models are insufficient to handle such situations.

A simple idea to account for multiple constraints consists in
taking the intersection of random graphs.
In this paper we initiate the study of random graphs so obtained
through a simple example. We examine the intersection of an 
Erd\H{o}s-R\'enyi graph and of one-dimensional geometric random graphs.
We investigate the zero-one laws for the 
property that there are no isolated nodes.
When the geometric component is defined on the unit circle, 
a full zero-one law is established and we determine its critical scaling. 
When the geometric component lies in the unit interval, there is a gap
in that the obtained zero and one laws are
found to express deviations from different critical scalings. 
In particular, the first moment method requires a 
larger critical scaling than in the unit circle case
in order to obtain the one law.
This discrepancy is somewhat surprising given that the zero-one laws 
for the absence of isolated nodes are identical in the geometric 
random graphs on both the unit interval and unit circle.
\end{abstract}


\section{Introduction}
\label{sec:Introduction}

Graphs provide simple and useful representations
for networks with the presence of an edge between a pair of nodes 
marking their ability to communicate with each other. 
Thus, for some set $V$ of nodes, an undirected 
graph $G \equiv (V,E)$ with edge set $E$ is defined such that an edge
exists between nodes $i$ and $j$ 
if and only if these nodes can establish a communication link. 
This adjacency between nodes in the graph 
representation may depend on various constraints, both physical
and logical. In typical settings, only a {\em single} adjacency constraint 
is considered. Here are some examples.

\begin{enumerate}
\item[(i)] In wireline networks, an edge between two nodes signifies the 
existence of a physical point-to-point communication link 
(e.g., fiber link) between the two nodes; 

\item[(ii)] Imagine a wireless network serving a set of
users distributed over a region $\mathbb{D}$ of the plane. 
A popular model, known as the disk model, postulates that
nodes $i$ and $j$ located at $\myvec{x}_i$ and $\myvec{x}_j$
in $\mathbb{D}$ are able to communicate if 
$\| \myvec{x}_i - \myvec{x}_j \| \leq r$ where $r$ is the
transmission range;

\item[(iii)] 
Eschenauer and Gligor \cite{EschenauerGligor} have recently proposed
a key pre-distribution scheme for use in wireless sensor networks:
Each node is randomly assigned a small set of distinct keys
from a large key pool. These keys form the key ring of the node,
and are inserted into its memory. Nodes 
can establish a secure link between them when they have at least one key 
in common in their key rings.
\end{enumerate}

Sometimes in applications there is a need to account 
for {\em multiple} adjacency constraints to reflect the several citeria
that must be satisfied before communication can take place between 
two users. For instance, consider the situation where
the Eschenauer-Gligor scheme is used in a wireless sensor network whose
nodes have only a finite transmission range (as is the case in practice).
Then, in order for a pair of nodes to establish a secure link, 
it is not enough that the distance between them 
does not exceed the transmission range.\footnote{This of course
assumes that the adopted communication model is compatible with
the disk model.}
They must also have at least one key in common.

Such situations  can be naturally formalized in the following setting: 
Suppose we have two adjacency constraints, say as in the example above,
modeled by the undirected graphs 
$G_1 \equiv (V,E_1)$ and $G_2 \equiv (V,E_2)$.
The {\em intersection} of these graphs is the graph 
$(V,E)$ with edge set $E$ given by
\[
E := E_1 \cap E_2 ,
\]
and we write $G_1 \cap G_2 := (V, E_1 \cap E_2 )$.
Through the intersection graph $G_1 \cap G_2$, we are 
able to simultaneously capture two different adjacency constraints. 
Of course the same approach can be extended to an arbitrary
number of constraints, but in the interest of concreteness
we shall restrict the discussion to the case of two constraints.

In an increasing number of contexts, 
{\em random} graph models\footnote{Here we shall 
understand random graphs in the broad sense to mean graph-valued random
variables.  To avoid any ambiguity with current usage,
the random graphs introduced by Erd\H{o}s and R\'enyi
in their groundbreaking paper \cite{ER1960} will be called
Erd\H{o}s-R\'enyi graphs.}
have been found to be more appropriate.
For instance, in wireless networking 
several classes of random graphs have been proposed
to model the effects of geometry, mobility and user interference
on the wireless communication link, e.g.,
geometric random graphs (also known as disk models)
\cite{GuptaKumar1998, HanMakowski-CommLetters, PenroseBook}
and signal-to-interference-plus-noise-ratio (SINR) graphs
\cite{DousseBaccelliThiran+Infocom2003,
DousseFranceschettiMacrisMeesterThiran}.
See Sections \ref{subsec:GRGs} and \ref{subsec:ERGs} 
for a description of the two classes of random graphs considered here.

When random graphs are used, we can also define their
intersection in an obvious manner:
Given two random graphs with vertex set $V$, say
$\mathbb{G}_1 \equiv (V,\mathbb{E}_1)$
and
$\mathbb{G}_2 \equiv (V,\mathbb{E}_2)$,
their intersection is the random graph
$(V,\mathbb{E})$ where
\[
\mathbb{E} := \mathbb{E}_1 \cap \mathbb{E}_2.
\]
For simplicity assume
the random graphs $\mathbb{G}_1$ and $\mathbb{G}_2$ to be independent. 
A natural question to ask is the following:
How are the structural properties
of the random graph $\mathbb{G}_1 \cap \mathbb{G}_2$
shaped by those of the random graphs $\mathbb{G}_1$ and $\mathbb{G}_2$?
Here we are particularly interested in zero-one laws for 
certain graph properties -- More on that later.

Intersecting graphs represents a modular approach
to building more complex models. It could be argued
that this approach is of interest only if the known
structural properties for the component random graphs can be leveraged
to gain a better understanding of the resulting
intersection graphs. 
As we shall see shortly through the simple example developed here, 
successfully completing this program
is not as straightforward as might have been expected.

In this paper we consider exclusively the random graph
obtained by intersecting Erd\H{o}s-R\'enyi graphs
with certain geometric random graphs in one dimension.
We were motivated to consider this simple model 
for the following reasons:

(i) The disk model popularized by the work of Gupta and Kumar
\cite{GuptaKumar1998}
assumes simplified pathloss, no user interference and no fading,
and the transmission range is a proxy for transmit
power to be used by the users. One crude way to include fading 
is to think of it as link outage.
Thus an edge is present between a pair
of nodes if and only if they are within communication range
(so that there is a communication link in the sense of
the usual disk model) {\em and}
that link between them is indeed active (i.e., not in outage).
This simple model is simply obtained by
taking the intersection of the disk model with
an Erd\H{o}s-R\'enyi graph.

(ii) Both Erd\H{o}s-R\'enyi graphs and geometric random graphs
are well understood classes of random graphs with an extensive literature
devoted to them; see the monographs
\cite{Bollobas, JansonLuczakRucinski, PenroseBook} for
Erd\H{o}s-R\'enyi graphs and the text \cite{PenroseBook} for
geometric random graphs. Additional information concerning
one-dimensional graphs can be found in the references
\cite{GodehardtBook}
\cite{GodehardtJaworski}
\cite{HanThesis}
\cite{HanMakowski-TransactionsIT+Uniform}
\cite{Maehara}.
It is hoped that this wealth of results 
will prove helpful in successfully carrying out the program outlined earlier.

(iii) Furthermore, this simple model is a trial balloon
for the study of more complicated situations. In particular, we have in mind
the study of wireless sensor networks employing the Eschenauer-Gligor
scheme to establish secure links.
In that case the resulting random intersection graphs,
the so-called random key graphs under partial visibility
\cite{DiPietroManciniMeiPanconesiRadhakrishnan2004}
\cite{DiPietroManciniMeiPanconesiRadhakrishnan2006}
\cite{YaganMakowskiISIT2008},
share some similarity with the models discussed here, but have far greater
complexity due to lack of independence in the link assignments in the
non-geometric component;
see comments in Section \ref{ConcludingRemarks}.
Our ability to successfully complete the study of the models
considered in this paper would provide  
some measure of comfort that the more complicated cases are indeed
amenable to analysis, with pointers to possible results.

We would like to draw attention to a similar problem which
has been studied recently. In \cite{YiWanLiFrieder}, the authors 
consider a geometric random graph where the {\em nodes} become inactive 
independently with a certain probability. In contrast, we are interested
in the situation where the {\em edges} in the geometric random graph 
can become inactive.

In the context of our simple model we investigate the zero-one laws 
for the property that there are no isolated nodes;
particular emphasis is put on identifying the corresponding
critical scalings. This is done with the help of 
the method of first and second moments.
Even this simple and well-structured situation gives rise 
to some surprising results:
When the geometric component is defined on the unit circle,
a full zero-one law is established and we determine its critical scaling. 
When the geometric component lies in the unit interval, there is a gap in
the results in that the obtained zero and one laws are
found to express deviations from different critical scalings. 
%The method of first and second moments is not 
%powerful enough to handle boundary effects. 
In particular, we encounter a 
situation where the first moment method requires a larger critical 
scaling than in the unit circle case in order to obtain the one law.
This discrepancy is somewhat surprising given that the zero-one laws 
for the absence of isolated nodes are identical in the geometric 
random graphs on both the unit interval and unit circle.
%This seems to lay the ground for the (perhaps naive)
Thus one is led to the (perhaps naive)
expectation that the boundary 
effects of the geometric component play no role in shaping 
the zero-one laws in the random intersection graphs.
%This discrepancy between the zero and one laws appears to be an artifact
%of the method of first moment, and a different approach is needed to bridge 
%this gap.
Therefore, it appears that this discrepancy between the zero and 
one laws is an artifact of the method of first moment, and a 
different approach is needed to bridge this gap.

The analysis given here provides some insight into classical results.
This is done by developing a new interpretation
of the critical scalings (for the absence of isolated nodes)
in terms of the probability of an edge existing
between a pair of nodes. This interpretation seems to hold quite generally.
In fact, it is this observation which enabled us to guess the form
of the zero-one law for the random intersection graphs and 
may find use in similar problems.

It is natural to wonder here what form take the zero-one laws for 
the property of graph connectivity. We remark that this is now
a more delicate problem for contrary to what occurs 
with one-dimensional graphs \cite{GodehardtBook} \cite{GodehardtJaworski}
\cite{HanThesis} \cite{HanMakowski-TransactionsIT+Uniform} \cite{Maehara},
the total ordering of the line cannot be used to advantage, and new
approaches are needed. But not all is lost:
In some sense the property that there are no isolated nodes can 
be viewed as a \lq\lq first-order approximation" to the 
property of graph connectivity -- This is borne out
by the fact that for many
classes of random graphs these two properties are asymptotically 
equivalent under the appropriate scaling; see
the monographs \cite{Bollobas} and \cite{PenroseBook} for
Erd\H{o}s-R\'enyi graphs and geometric random graphs, respectively.
In that sense the preliminary results obtained here
constitute a first step on the road to
establish zero-one laws for the property of graph connectivity. 

A word on the notation and conventions in use:
Throughout $n$ will denote the number of nodes in the random graph and 
all limiting statements, including asymptotic equivalences, are understood
with $n$ going to infinity.
The random variables (rvs) under consideration are all
defined on the same probability triple $(\Omega, {\cal F}, \mathbb{P})$.
Probabilistic statements are made with respect to this probability
measure $\mathbb{P}$, and we denote the corresponding expectation
operator by $\mathbb{E}$.
Also, we use the notation $=_{st}$ to indicate distributional equality. 
The indicator function of an event $E$ is denoted by $\1{E}$.

\section{Model and assumptions}
\label{sec:ModelAssumptions}

In this paper we are only concerned with undirected graphs.
As usual, a graph $G \equiv (V,E)$ is said to be {\em connected} if every 
pair of nodes in $V$ can be linked by at least one path over the edges
(in $E$) of the graph. We say a node is {\em isolated} if no edge exists 
between the node and any of the remaining nodes. 
Also, let $\mathcal{E}(G)$ refer to the set of edges of $G$, namely 
$\mathcal{E}(G) = E$. We begin by recalling the classical random graph models 
used in the definition of the model analyzed here.

\subsection{The geometric random graphs}
\label{subsec:GRGs}

Two related geometric random graphs are introduced.
Fix $n =2, 3, \ldots $ and $r > 0$,
and consider a collection $X_1, \ldots , X_n$ of i.i.d. rvs
which are distributed uniformly in the interval $[0,1]$ (referred
to as the {\em unit interval}).
We think of $r$ as the transmission range and $X_1, \ldots , X_n$ 
as the locations of $n$ nodes (or users), labelled $1, \ldots , n$, 
in the interval $[0,1]$.

Nodes $i$ and $j$ are said to be adjacent if 
$|X_i - X_j  | \leq r $, 
in which case an undirected edge exists between them.
The indicator rv $\chi^{(L)}_{ij} (r)$ that
nodes $i$ and $j$ are adjacent is given by
\[
\chi^{(L)}_{ij} (r)
:= \1{ |X_i -X_j| \leq r } .
\]
This notion of edge connectivity gives rise to
an undirected geometric random graph on the unit interval,
thereafter denoted $\mathbb{G}^{(L)} (n; r)$.

For each $i=1, \ldots , n$, node $i$ is 
an isolated node in $\mathbb{G}^{(L)} (n; r)$ if
$|X_i - X_j| > r$ for all $j=1, \ldots , n$ with $j\neq i$.
The indicator rv $\chi^{(L)}_{n,i}(r)$
that node $i$ is an isolated node in $\mathbb{G}^{(L)} (n; r)$ is
given by
\[
\chi^{(L)}_{n,i}(r)
:= \prod_{j=1, j \neq i}^n 
\left ( 1 - \chi^{(L)}_{ij} (r) \right ).
\]
The number of isolated nodes in $\mathbb{G}^{(L)} (n; r)$ 
is then given by
\[
I^{(L)}_n (r) := \sum_{i=1}^n \chi^{(L)}_{n,i}(r).
\]

We also consider the geometric random graph obtained by
locating the nodes uniformly on the circle with unit circumference
(thereafter referred to as the {\em unit circle}) -- This corresponds
to identifying the end points of the unit interval.
In this formulation, we fix some reference point on the circle 
and the node locations $X_1,\dots,X_n$ are given by the length 
of the clockwise arc with respect to this reference point. 
We measure the distance between any two 
nodes by the length of the smallest arc between the nodes,
i.e., the distance between nodes $i$ and $j$ is given by
\[
\| X_i-X_j \|
:=\min(|X_i-X_j|,1-|X_i-X_j|).
\]
As we still think of $r$ as the transmission range,
nodes $i$ and $j$ are now said to be adjacent if 
$ \| X_i-X_j \| \leq r.$ 
The indicator rv $\chi^{(C)}_{ij} (r)$ that
nodes $i$ and $j$ are adjacent is given by
\[
\chi^{(C)}_{ij} (r)
:= \1{ \|X_i-X_j \| \leq r }.
\]
This notion of adjacency leads to
an undirected geometric random graph on the unit circle,
thereafter denoted $\mathbb{G}^{(C)} (n; r)$.
This model is simpler to 
analyze as the boundary effects have been removed. 

The indicator rv $\chi^{(C)}_{n,i}(r)$ that node $i$ is an isolated node
in $\mathbb{G}^{(C)} (n; r)$ is again defined by
\[
\chi^{(C)}_{n,i}(r)
:= \prod_{j=1, j \neq i}^n 
\left ( 1 - \chi^{(C)}_{ij} (r) \right ).
\]
The number of isolated nodes in $\mathbb{G}^{(C)} (n; r)$ 
is then given by
\[
I^{(C)}_n (r) := \sum_{i=1}^n \chi^{(C)}_{n,i}(r).
\]

Throughout, it will be convenient to view the graphs
$\mathbb{G}^{(L)} (n; r)$ and $\mathbb{G}^{(C)} (n; r)$ 
as {\em coupled} in that they are constructed from the {\em same}
rvs $X_1, \ldots , X_n$ defined on the {\em same}
probability space $(\Omega, {\cal F}, \mathbb{P})$.

Note that the two models differ only in the manner in which the 
distance between two users is defined. To take advantage 
of this observation, we shall write
\[
d(x,y) :=
\left \{
\begin{array}{ll}
|x-y| & \mbox{on~the~unit~interval} \\
  &                 \\
\| x-y \| 
& \mbox{on~the~unit~circle} 
\end{array}
\right .
\]
for all $x,y$ in $[0,1]$ as a compact way to capture the appropriate
notion of \lq\lq distance".
Also, in the same spirit, as a way to lighten the notation,
we omit the superscripts $(L)$ 
and $(C)$ from the notation when the discussion 
applies equally well to both cases.

\subsection{The Erd\H{o}s-R\'enyi graphs}
\label{subsec:ERGs}

Fix $n=2,3, \ldots $ and $p$ in $[0,1]$. In this case, $p$ corresponds
to the probability that an (undirected) edge exists between any pair of nodes.
We start with rvs $\{ B_{ij}(p), \ 1 \leq i < j \leq n \},$ which are i.i.d. 
$\{0,1 \}$-valued rvs with success probability $p$.
Nodes $i$ and $j$ are said to be adjacent if $B_{ij} (p) = 1$.
This notion of edge connectivity defines
the undirected Erd\H{o}s-R\'enyi (ER) random graph,
thereafter denoted $\mathbb{G} (n; p)$.

For each $i=1, \ldots , n$, node $i$ is isolated in
$\mathbb{G} (n; p)$ if 
$B_{ij}(p) = 0$ for $i < j \leq n $ and
$B_{ji}(p) = 0$ for $1\leq j < i  $. The indicator $\chi_{n,i}(p)$
that node $i$ is an isolated node in $\mathbb{G} (n;p)$ is
then given by
\[
\chi_{n,i}(p)
:= \prod_{i < j \leq n } \left ( 1 - B_{ij}(p) \right ) 
\cdot \prod_{1\leq j < i } \left ( 1 - B_{ji} (p) \right ).
\]
The number of isolated nodes in $\mathbb{G} (n;p)$ is the rv $I_n(p)$
given by
\[
I_n (p) := \sum_{i=1}^n \chi_{n,i}(p).
\]

\subsection{Intersecting the geometric and Erd\H{o}s-R\'enyi graphs}
\label{subsec:Intersection}

The random graph model studied in this paper
is parametrized by the number $n$ of nodes, the transmission
range $r > 0$ and the probability $p$ ($ 0 \leq p \leq 1 $)
that a link is active (i.e., not in outage). 
To lighten the notation we often group the parameters $r$ and $p$ 
into the ordered pair $\myvec{\theta} \equiv (r,p)$.

Throughout we always assume that the collections of rvs
$\{ X_i, \ i=1, \ldots , n \}$ and
$\{ B_{ij}(p), \ 1 \leq i < j \leq n \}$ are {\em mutually independent}.
With the convention introduced earlier,
the intersection of the two graphs 
$\mathbb{G} (n;r)$ and $\mathbb{G} (n;p)$ is the graph
\[
\mathbb{G}(n;\myvec{\myvec{\theta}})
:=
\mathbb{G} (n; r)\cap \mathbb{G} (n;p)
\]
defined on the vertex set $\{ 1, \ldots , n\}$ with edge set
given by
\[
\mathcal{E} \left ( \mathbb{G}(n;\myvec{\theta}) \right )
=
\mathcal{E} \left ( \mathbb{G} (n; r) \right )
\cap \mathcal{E} \left ( \mathbb{G} (n;p) \right ).
\]
We refer to $\mathbb{G}(n;\myvec{\theta})$ as the intersection graph
on the unit interval (resp. unit circle)
when in this definition, $\mathbb{G} (n; r)$ is taken to be
$\mathbb{G}^{(L)}(n; r)$ (resp.  $\mathbb{G}^{(C)}(n; r)$).

The nodes $i$ and $j$ are adjacent in
$\mathbb{G}(n;\myvec{\theta})$ if and only if 
they are adjacent in {\em both}
$\mathbb{G} (n; r)$ {\rm and} $\mathbb{G} (n;p)$.
The indicator rv $\chi_{ij} (\myvec{\theta})$ that
nodes $i$ and $j$ are adjacent in 
$\mathbb{G}(n;\myvec{\theta})$ is given by
\[
\chi_{ij} (\myvec{\theta})
=
\left \{
\begin{array}{ll}
\chi_{ij} (r) B_{ij}(p)  & \mbox{if~ $ i < j $} \\
  &                 \\
\chi_{ij} (r) B_{ji}(p)  & \mbox{if~ $ j < i $.} 
\end{array}
\right .
\]

For each $i=1, \ldots , n$, node $i$ is isolated 
in $\mathbb{G}(n;\myvec{\theta})$
if either it is not within transmission range from
each of the $(n-1)$ remaining nodes, or being within range
from some nodes, the corresponding links all are inactive.
The indicator rv $\chi_{n,i}(\myvec{\theta})$ that node $i$ is an 
isolated node in $\mathbb{G} (n;\myvec{\theta})$ can be expressed as
\[
\chi_{n,i}(\myvec{\theta})
:= \prod_{i=1,\ j \neq i}^n \left ( 1 - \chi_{ij} (\myvec{\theta}) \right ) .
\]
As expected, the number of isolated nodes in $\mathbb{G} (n;\myvec{\theta})$ 
is similarly defined as
\[
I_n (\myvec{\theta}) := \sum_{i=1}^n \chi_{n,i}(\myvec{\theta}).
\]

\subsection{Scalings}

Some terminology:
A scaling for either of the geometric graphs is a mapping
$r: \mathbb{N}_0 \rightarrow \mathbb{R}_+$, while
a scaling for ER graphs is simply a mapping
$p: \mathbb{N}_0 \rightarrow [0,1]$.
A scaling for the intersection graphs 
combines scalings for each of the component graphs, and
is defined as a mapping
$\myvec{\theta}: \mathbb{N}_0 \rightarrow \mathbb{R}_+ \times [0,1]$.

The main objective of this paper can be stated as follows:
Given that
\[
\bP{ \mathbb{G}(n;\myvec{\theta}_n) ~\mbox{\rm has~no~isolated~nodes} }
= \bP{ I_n(\myvec{\theta}_n)=0 }
\]
for all $n=2,3, \ldots $, what conditions are needed on the scaling 
$\myvec{\theta}: \mathbb{N}_0 \rightarrow \mathbb{R}_+ \times [0,1]$
to ensure that
\[
\lim_{n \rightarrow \infty}
\bP{ I_n(\myvec{\theta}_n)=0 } = 1 
\quad (\mbox{\rm resp.} ~ 0).
\]
In the literature such results are known as
zero-one laws. Interest in them stems from their ability to
capture the threshold behavior of the underlying random graphs.

\section{Classical results}
\label{sec:ClassicalResultsER+GRG}

\subsection{Erd\H{o}s-R\'enyi graphs}
\label{subsec:ErdosrenyiGraphs}

There is no loss of generality in writing
a scaling $p: \mathbb{N}_0 \rightarrow [0,1]$ in the form
\begin{equation}
p_n = \frac{ \log n + \alpha_n }{n},
\quad n=1,2, \ldots
\label{eq:DeviationCondition+ER}
\end{equation}
for some deviation function
$\alpha: \mathbb{N}_0 \rightarrow \mathbb{R}$.
The following result is well known \cite{Bollobas, JansonLuczakRucinski}.

\begin{theorem}
{\sl For any scaling $p: \mathbb{N}_0 \rightarrow [0,1]$ in the form
(\ref{eq:DeviationCondition+ER}), we have the zero-one law
\begin{equation*}
\lim_{n \rightarrow \infty } 
\bP{ I_n(p_n)=0 } 
= \left \{
\begin{array}{ll}
0 & \mbox{if~ $\lim_{ n\rightarrow \infty }\alpha_n = - \infty $} \\
  &                 \\
1 & \mbox{if~ $\lim_{ n\rightarrow \infty }\alpha_n = + \infty $}
\end{array}
\right .
\label{eq:VeryStrongZeroOneLaw+NoIsolatedNodes+ER}
\end{equation*}
where the deviation function $\alpha : \mathbb{N}_0 \rightarrow \mathbb{R}$
is determined through (\ref{eq:DeviationCondition+ER}).
}
\label{thm:VeryStrongZeroOneLaw+NoIsolatedNodes+ER}
\end{theorem}

This result identifies the scaling
$p^\star: \mathbb{N}_0 \rightarrow [0,1]$ given by
\begin{equation*}
p^\star_n = \frac{\log n}{n},
\quad n=1,2, \ldots
\label{eq:CriticalScaling+ER}
\end{equation*}
as the critical scaling for the absence of isolated nodes in ER graphs.

\subsection{Geometric random graphs}
\label{subsec:GeometricRandomGraphs}

Any scaling $r: \mathbb{N}_0 \rightarrow \mathbb{R}_+$
can be written in the form
\begin{equation} 
r_n = \frac{\log n + \alpha_n}{2n},
\quad n=1,2, \ldots
\label{eq:TauScaling}
\end{equation}
for some deviation function
$\alpha: \mathbb{N}_0 \rightarrow \mathbb{R}$.
The following result can be found in 
\cite{AppelRusso}, \cite{PenroseBook}.

\begin{theorem}
{\sl For any scaling $r: \mathbb{N}_0 \rightarrow \mathbb{R}_+$ 
written in the form (\ref{eq:TauScaling}) for some deviation function
$\alpha: \mathbb{N}_0 \rightarrow \mathbb{R}$,
we have the zero-one law
\begin{equation*}
\lim_{n \rightarrow \infty } 
\bP{ I_n(r_n)=0 } 
= \left \{
\begin{array}{ll}
0 & \mbox{if~ $\lim_{ n\rightarrow \infty }\alpha_n = - \infty $} \\
  &                 \\
1 & \mbox{if~ $\lim_{ n\rightarrow \infty }\alpha_n = + \infty $}.
\end{array}
\right .
\label{eq:VeryStrongZeroOneLaw+Iso+Geo}
\end{equation*}
}
\label{thm:VeryStrongZeroOneLaw+Iso+Geo}
\end{theorem}

Theorem \ref{thm:VeryStrongZeroOneLaw+Iso+Geo}
identifies the scaling
$r^\star: \mathbb{N}_0 \rightarrow [0,1]$ given by
\begin{equation}
r^\star_n = \frac{\log n }{2n},
\quad n=1,2, \ldots
\label{eq:CriticalScaling+GRG}
\end{equation}
as the critical scaling for the absence of isolated nodes 
in geometric random graphs.

For reasons that will become apparent shortly, we now develop
an equivalent version of Theorem \ref{thm:VeryStrongZeroOneLaw+Iso+Geo}
that bears a striking resemblance with the zero-one law of
Theorem \ref{thm:VeryStrongZeroOneLaw+NoIsolatedNodes+ER} for ER graphs.

To that end, define
\[
\ell (r):=\min(1,2r),
\quad r \geq 0.
\]
Intuitively, $\ell(r)$ is akin to the probability that an edge exists 
between any pair of nodes in $\mathbb{G}(n;r)$ -- In fact it has exactly that
meaning for $\mathbb{G}^{(C)}(n;r)$ while it is true approximately (when
boundary conditions are ignored) for $\mathbb{G}^{(L)}(n;r)$.

With this in mind, for any scaling  
$r: \mathbb{N}_0 \rightarrow \mathbb{R}_+$ write
\begin{equation}
\ell(r_n)
= \frac{ \log n + \beta_n}{n},
\quad n=1,2, \ldots
\label{eq:TauScaling+Alternative}
\end{equation}
for some deviation function
$\beta: \mathbb{N}_0 \rightarrow \mathbb{R}$.
The representations (\ref{eq:TauScaling}) 
and (\ref{eq:TauScaling+Alternative}) together require
\[
\beta_n = \min(\alpha_n, n-\log n), \quad n=1,2, \ldots
\]
It is easily verified that 
$\lim_{n \rightarrow \infty} \beta_n = -\infty$
(resp. $\lim_{n \rightarrow \infty} \beta_n = +\infty$)
if and only if $\lim_{n \rightarrow \infty} \alpha_n = -\infty$
(resp. $\lim_{n \rightarrow \infty} \alpha_n = +\infty$).
This implies the following equivalent rephrasing of
Theorem \ref{thm:VeryStrongZeroOneLaw+Iso+Geo}.

\begin{theorem} 
{\sl For any scaling $r: \mathbb{N}_0 \rightarrow \mathbb{R}_+$
written in the form (\ref{eq:TauScaling+Alternative}) 
for some deviation function $\beta: \mathbb{N}_0 \rightarrow \mathbb{R}$,
we have the zero-one law
\begin{equation*}
\lim_{n \rightarrow \infty }
\bP{ I_n(r_n)=0 }
= \left \{
\begin{array}{ll}
0 & \mbox{if~ $\lim_{ n\rightarrow \infty }\beta_n = - \infty $} \\
  &                 \\
1 & \mbox{if~ $\lim_{ n\rightarrow \infty }\beta_n = + \infty $}.
\end{array}
\right .
\label{eq:VeryStrongZeroOneLaw+Iso+Geo+Alternative}
\end{equation*}
}
\label{thm:VeryStrongZeroOneLaw+Iso+Geo+Alternative}
\end{theorem}

By Theorem \ref{thm:VeryStrongZeroOneLaw+Iso+Geo+Alternative}
any scaling $r^\star: \mathbb{N}_0 \rightarrow \mathbb{R}_+$ 
such that
\begin{equation}
\ell(r^\star_n) = \frac{\log n }{n}, 
\quad n=1,2,\ldots 
\label{eq:CriticalScaling+GRG+Alternative}
\end{equation}
is a critical scaling for the absence of isolated nodes 
in geometric random graphs. 
As expected, it is easy to see that any scaling is critical under 
the definition (\ref{eq:CriticalScaling+GRG+Alternative})
if and only if it is under
the definition (\ref{eq:CriticalScaling+GRG}).

\section{The basic difficulty}

\subsection{Intersecting Erd\H{o}s-R\'enyi graphs}
\label{subsec:IntersectingER}

As a d\'etour consider intersecting two {\em independent} ER graphs.
This results in another ER graph, i.e.,
\[
\mathbb{G}(n;p) \cap \mathbb{G}(n;p^\prime ) 
=_{st}
\mathbb{G}(n;pp^\prime ) ,
\quad 0 \leq p,p^\prime \leq 1 .
\]
It is therefore a simple matter to select scalings
$p,p^\prime: \mathbb{N}_0 \rightarrow [0,1]$ such that the
intersection graph
$\mathbb{G}(n;p) \cap \mathbb{G}(n;p^\prime )$ exhibits a zero-one law
for the absence of isolated nodes. By Theorem 
\ref{thm:VeryStrongZeroOneLaw+NoIsolatedNodes+ER},
it suffices to take these scalings such that
\begin{equation*}
p_n p^\prime_n
= \frac{ \log n + \alpha_n }{n},
\quad n=1,2, \ldots
\end{equation*}
for some appropriate deviation function
$\alpha: \mathbb{N}_0 \rightarrow \mathbb{R}$.

Despite its simplicity, this result has some interesting implications:
For instance, select the two scalings such that
\[
p_n p^\prime_n
= \frac{1}{2} \frac{ \log n }{n},
\quad n=1,2, \ldots
\]
with
\[
p_n = p^\prime_n
= \sqrt{ \frac{1}{2} \frac{ \log n }{n}},
\quad n=1,2, \ldots
\]
In that case, upon writing
\[
\frac{1}{2} \frac{ \log n }{n}
= \frac{ \log n + \left ( -\frac{1}{2} \log n \right ) }{n},
\quad n=1,2, \ldots ,
\]
we conclude
\[
\lim_{n \rightarrow \infty} 
\bP{ \mathbb{G}(n;p_n) \cap \mathbb{G}(n;p^\prime_n )
     ~\mbox{\rm has~no~isolated~nodes} } = 0 
\]
by the zero law of Theorem 
\ref{thm:VeryStrongZeroOneLaw+NoIsolatedNodes+ER}.
Yet, we also have 
\[
\lim_{n \rightarrow \infty} 
\bP{ \mathbb{G}(n;p_n) ~\mbox{\rm has~no~isolated~nodes} }
=1
\]
and
\[
\lim_{n \rightarrow \infty} 
\bP{ \mathbb{G}(n;p^\prime_n) ~\mbox{\rm has~no~isolated~nodes} }
= 1
\]
by the one law of Theorem
\ref{thm:VeryStrongZeroOneLaw+NoIsolatedNodes+ER}
as we note that
\[
\sqrt{ \frac{1}{2} \frac{ \log n }{n}}
= \frac{ \log n + \alpha_n }{n}
\]
for all $n=1,2, \ldots$ with the choice
\[
\alpha_n := \sqrt{ \frac{n \log n}{2} } - \log n .
\]

Thus, even when the individual graphs 
$\mathbb{G}(n;p_n )$ and $\mathbb{G}(n;p^\prime_n )$ 
contain {\em no} isolated nodes with a probability close to one, 
it is possible for the 
intersection graph $\mathbb{G}(n;p_n) \cap \mathbb{G}(n;p^\prime_n )$ 
to contain isolated nodes with a probability very close to one.
The reason for this is quite simple: A node that is isolated
in $\mathbb{G}(n;p_np_n^\prime )$
may not be isolated in either of the component graphs
$\mathbb{G}(n;p_n)$ and $\mathbb{G}(n;p_n^\prime )$.
For ER graphs, the answer, although very simple, 
fails to give much insight into how the individual graphs interact 
with each other and how this affects
the overall behavior of the intersection graph.

\subsection{Intersecting an Erd\H{o}s-R\'enyi graph 
            with a geometric random graph}

With this in mind, note that with $0 < r < 1$ for the unit interval 
(resp. $0 < r < 0.5$ for the unit circle) and $0 < p < 1$,
the intersection graph
$\mathbb{G}(n;r) \cap \mathbb{G}(n;p)$ is {\em not}
stochastically equivalent 
to either a geometric random graph or an Erd\H{o}s-R\'enyi graph, i.e.,
it is {\em not} possible to find parameters
$r^\prime = r^\prime(n;r,p)$ and $p^\prime = p^\prime(n;r,p)$
in $\mathbb{R}_+$ and $[0,1]$, respectively, such that
\[
\mathbb{G}(n;r) \cap \mathbb{G}(n;p)
=_{st} 
\mathbb{G}(n;r^\prime)
\]
and
\[
\mathbb{G}(n;r) \cap \mathbb{G}(n;p)
=_{st} 
\mathbb{G}(n;p^\prime ) .
\]
Consequently, results for either ER or geometric random graphs
(as given in Section \ref{sec:ClassicalResultsER+GRG}) cannot be used 
in a straightforward manner to determine the zero-one laws for 
the intersection graphs.

On the other hand, it is obvious that
if either $\mathbb{G}(n;r)$ or $\mathbb{G}(n;p)$ contains isolated nodes,
then 
$\mathbb{G}(n;r) \cap \mathbb{G}(n;p)$ must contain isolated nodes.
Therefore, a zero law for the intersection graph should 
follow by combining the zero laws for the ER and
geometric random graphs. However, as will become apparent from our main results,
such arguments are too loose to provide the best possible zero law.

A direct approach is therefore required with the difficulty 
mentioned earlier remaining, namely that a node isolated
in $\mathbb{G}(n;r) \cap \mathbb{G}(n;p)$
may not be isolated in either $\mathbb{G}(n;r)$ or $\mathbb{G}(n;p)$.
Nevertheless the corresponding zero-one laws do provide a basis 
for guessing the form of the zero-one law for the intersection graphs.
This is taken on in the next section.

\section{The main results}
\label{sec:MainResults}

\subsection{Guessing the form of the results}
\label{subsec:Guessing}

Upon comparing the zero-one laws of
Theorems \ref{thm:VeryStrongZeroOneLaw+NoIsolatedNodes+ER}
and \ref{thm:VeryStrongZeroOneLaw+Iso+Geo+Alternative},
the following shared structure suggests itself:
For the random graphs of interest here (as well as for others,
e.g., random key graphs \cite{YaganMakowskiISIT2008}), 
it is possible to identify a quantity which gives 
\[
\bP{\textrm{Edge exists between two nodes}} ,
\]
either exactly (e.g., $p$ in ER graphs or $\ell(r)$ in geometric random
graphs on the circle) or approximately (e.g., $\ell(r)$ in geometric
random graphs on the interval).
Critical scalings for the absence of isolated nodes are then determined
through the requirement
\begin{equation}
\bP{\textrm{Edge exists between two nodes}} 
= \frac{\log n}{n}.
\label{eq:CriticalRequirement}
\end{equation}
In particular the zero-one law requires scalings satisfying
\[
\begin{array}{c}
\bP{\textrm{Edge exists between two nodes}} \\
\mbox{\rm \lq\lq much~smaller\rq\rq ~than~} \frac{\log n}{n},
\end{array}
\]
while the one-one law deals with scalings satisfying
\[
\begin{array}{c}
\bP{\textrm{Edge exists between two nodes}} \\
\mbox{\rm \lq\lq much~larger\rq\rq ~than~} \frac{\log n}{n}.
\end{array}
\]
The exact technical meaning of \lq\lq much smaller" and
\lq\lq much larger" forms the content of results such as
Theorems \ref{thm:VeryStrongZeroOneLaw+NoIsolatedNodes+ER}
and \ref{thm:VeryStrongZeroOneLaw+Iso+Geo+Alternative}.

With this in mind, for the random intersection graphs studied here
it is natural to take
\begin{equation}
\bP{\textrm{Edge exists between two nodes}} := p \ell(r)
\label{eq:CriticalRequirement2}
\end{equation}
under the enforced independence assumptions.
We expect that a critical scaling 
$\myvec{\theta}^\star : \mathbb{N}_0 \rightarrow \mathbb{R}_+ \times [0,1]$
for the random intersection graphs should be determined by
\begin{equation*}
p^\star_n \ell(r^\star_n)  = \frac{\log n}{n},
\quad n=1,2, \ldots
\label{eq:CriticalScaling+Intersection}
\end{equation*}
The exact form taken by the results is discussed in 
Sections \ref{subsec:IntersectionGraphsUnitCircle}
and \ref{subsec:IntersectionGraphsUnitInterval}.
We start with the model on the circle
for which we have obtained the most complete results.

\subsection{Intersection graphs on the unit circle}
\label{subsec:IntersectionGraphsUnitCircle}

With a scaling 
$\myvec{\theta}: \mathbb{N}_0 \rightarrow \mathbb{R}_+ \times [0,1]$,
we associate the sequence $\alpha: \mathbb{N}_0 \rightarrow \mathbb{R}$
through 
\begin{equation}
p_n \ell (r_n)  =\frac{\log n + \alpha_n}{n},
\quad n=1,2, \ldots 
\label{eq:DeviationUnitCircle}
\end{equation}
In the case of the intersection graph on the unit circle
we get a full zero-one law.

\begin{theorem}[Unit circle]
{\sl For any
scaling $\myvec{\theta}: \mathbb{N}_0 \rightarrow \mathbb{R}_+ \times [0,1]$,
we have the zero-one law
\begin{equation*}
\lim_{n \rightarrow \infty } 
\bP{ I^{(C)}_n(\myvec{\theta}_n)=0 } 
= \left \{
\begin{array}{ll}
0 & \mbox{if~ $\lim_{ n\rightarrow \infty }\alpha_n = - \infty $} \\
  &                 \\
1 & \mbox{if~ $\lim_{ n\rightarrow \infty }\alpha_n = + \infty $}
\end{array}
\right .
\label{eq:VeryStrongZeroOneLaw+Iso+Int1}
\end{equation*}
where the sequence $\alpha : \mathbb{N}_0 \rightarrow \mathbb{R}$
is determined through (\ref{eq:DeviationUnitCircle}).
}
\label{thm:MainThmCir}
\end{theorem}

\subsection{Intersection graphs on the unit interval}
\label{subsec:IntersectionGraphsUnitInterval}

With a scaling 
$\myvec{\theta}: \mathbb{N}_0 \rightarrow \mathbb{R}_+ \times [0,1]$,
we also associate the sequence 
$\alpha^\prime: \mathbb{N}_0 \rightarrow \mathbb{R}_+ $
through
\begin{equation}  
p_n \ell (r_n)  =\frac{2(\log n - \log \log n ) + \alpha^\prime_n}{n},
\quad n=1,2, \ldots
\label{eq:DeviationUnitInterval}
\end{equation}
For the intersection graph on the unit interval 
there is a gap between the zero and one laws.

\begin{theorem}[Unit interval]
{\sl For any scaling 
$\myvec{\theta}: \mathbb{N}_0 \rightarrow \mathbb{R}_+ \times [0,1]$,
we have the zero-one law
\begin{equation*}
\lim_{n \rightarrow \infty } 
\bP{ I^{(L)}_n(\myvec{\theta}_n)=0 } 
= \left \{
\begin{array}{ll}
0 & \mbox{if~ $\lim_{ n\rightarrow \infty }\alpha_n = - \infty $} \\
  &                 \\
1 & \mbox{if~ $\lim_{ n\rightarrow \infty }\alpha^\prime_n = + \infty $}
\end{array}
\right .
\label{eq:VeryStrongZeroOneLaw+Iso+Int2}
\end{equation*}
where the sequences $\alpha, \alpha^\prime : 
\mathbb{N}_0 \rightarrow \mathbb{R}$
are determined through (\ref{eq:DeviationUnitCircle}) 
and (\ref{eq:DeviationUnitInterval}), respectively.
}
\label{thm:MainThmInt}
\end{theorem}

An elementary coupling argument shows that
for any particular realization of the rvs $\{ X_i, \ i=1, \ldots , n \}$ 
and $\{ B_{ij}(p), \ 1 \leq i < j \leq n \}$,
the graph on the circle contains 
more edges than the graph on the interval.
As a result, the zero law for the unit circle automatically implies 
the zero law for the unit interval, and only the former needs to be
established.

\section{Method of first and second moments}
\label{sec:FirstAndSecond}

The proofs rely on the method of first and second moments
\cite[p. 55]{JansonLuczakRucinski}, 
an approach widely used in the theory of Erd\H{o}s-R\'enyi graphs:
Let $Z$ denote an $\mathbb{N}$-valued rv with finite second moment.
The method of first moment
\cite[Eqn. (3.10), p. 55]{JansonLuczakRucinski}
relies on the inequality
\begin{equation}
1 - \bE{ Z } \leq \bP{ Z = 0 },
\label{eq:LowerBoundFunction}
\end{equation}
while the method of second moment
\cite[Remark 3.1, p. 55]{JansonLuczakRucinski}
uses the bound
\begin{equation}
\bP{ Z = 0 }
\leq
1 - \frac{ \bE{Z}^2}{\bE{Z ^2} } .
\label{eq:UpperBoundFunction}
\end{equation}

Now, pick a scaling 
$\myvec{\theta}: \mathbb{N}_0 \rightarrow \mathbb{R}_+ \times [0,1]$.
From (\ref{eq:LowerBoundFunction}) we see that the one law
\begin{equation*}
\lim_{n \rightarrow \infty} \bP{ I_n(\myvec{\theta}_n) = 0 } = 1
\label{eq:GenericOneLaw}
\end{equation*}
is established if we show that
\begin{equation}
\lim_{n \rightarrow \infty} \bE{ I_n(\myvec{\theta}_n) } = 0.
\label{eq:GenericOneLaw+Condition}
\end{equation}
%if $\lim_{n \rightarrow \infty} \alpha_n = + \infty$ or
%$\lim_{n \rightarrow \infty} \alpha'_n = + \infty$ as appropriate.
On the other hand, it is plain from (\ref{eq:UpperBoundFunction}) that
\begin{equation*}
\lim_{n \rightarrow \infty} \bP{ I_n(\myvec{\theta}_n) = 0 } = 0
\label{eq:GenericZeroLaw}
\end{equation*}
if 
\begin{equation}
\liminf_{n \rightarrow \infty} 
\left ( \frac{ \bE{ I_n (\myvec{\theta}_n) }^2 }{ \bE{ I_n (\myvec{\theta}_n)^2 } }
\right )
\geq 1 .
\label{eq:GenericZeroLaw+Condition}
\end{equation}

Upon using the exchangeability and the binary nature of the rvs involved 
in the count variables of interest, 
we can obtain simpler characterizations of the convergence statements
(\ref{eq:GenericOneLaw+Condition}) and (\ref{eq:GenericZeroLaw+Condition}). 
Indeed, for all $n=2,3, \ldots $ and  every $\myvec{\theta}$ in 
$\mathbb{R}_+ \times [0,1]$, the calculations 
\[
\bE{ I_n (\myvec{\theta}) }
= \sum_{i=1}^n \bE{ \chi_{n,i}(\myvec{\theta}) }
= n \bE{ \chi_{n,1}(\myvec{\theta}) }
\]
and
\begin{eqnarray}
& & \bE{ I_n(\myvec{\theta})^2 }
\nonumber \\
&=& 
\bE{\left (\sum_{i=1}^n \chi_{n,i}(\myvec{\theta}) \right )^2 }
\nonumber \\
&=& \sum_{i=1}^n \bE{ \chi_{n,i}(\myvec{\theta}) }
+ \sum_{i,j=1, \ i\neq j }^n \bE{ \chi_{n,i}(\myvec{\theta}) \chi_{n,j} (\myvec{\theta}) }
\nonumber \\
&=& n \bE{ \chi_{n,1}(\myvec{\theta}) }
+ n(n-1) \bE{ \chi_{n,1}(\myvec{\theta}) \chi_{n,2} (\myvec{\theta}) }
\nonumber
\end{eqnarray}
are straightforward, so that
\begin{eqnarray}
& & \frac{ \bE{ I_n(\myvec{\theta})^2 } }{ \bE{ I_n(\myvec{\theta}) }^2 } 
\nonumber \\
&=&
\frac{ 1 }{ n \bE{ \chi_{n,1}(\myvec{\theta}) } }
+ \frac{n-1}{n}
\cdot
\frac{ \bE{ \chi_{n,1}(\myvec{\theta}) \chi_{n,2} (\myvec{\theta}) } }
       { \bE{ \chi_{n,1} (\myvec{\theta}) }^2 } .
\nonumber
\end{eqnarray}

Thus, for the given scaling 
$\myvec{\theta}: \mathbb{N}_0 \rightarrow \mathbb{R}_+ \times [0,1]$,
we obtain the one law by showing that
\begin{equation}
\lim_{n \rightarrow \infty}
n \bE{ \chi_{n,1}(\myvec{\theta}_n) } = 0, 
\label{eq:ToShow1}
\end{equation}
while the zero law will follow if we show that
\begin{equation}
\lim_{n \rightarrow \infty} n \bE{ \chi_{n,1}(\myvec{\theta}_n) } 
= \infty
\label{eq:ToShow2}
\end{equation}
and
\begin{equation}
\limsup_{n \rightarrow \infty} 
\left (
\frac{ \bE{ \chi_{n,1}(\myvec{\theta}_n) \chi_{n,2} (\myvec{\theta}_n) } }
       { \bE{ \chi_{n,1} (\myvec{\theta}_n) }^2 }
\right )
\leq 1.
\label{eq:ToShow3}
\end{equation}
The bulk of the technical discussion therefore amounts to establishing 
(\ref{eq:ToShow1}), (\ref{eq:ToShow2}) and (\ref{eq:ToShow3}) 
under the appropriate conditions on the scaling 
$\myvec{\theta} : \mathbb{N}_0 \rightarrow \mathbb{R}_+ \times [0,1]$.

To that end, in the next two sections 
we derive expressions for the quantities
entering (\ref{eq:ToShow1}), (\ref{eq:ToShow2}) and (\ref{eq:ToShow3}).
Throughout we denote by $X$, $Y$ and $Z$ three mutually independent
rvs which are uniformly distributed on $[0,1]$, and by $B$, $B^\prime$ 
and $B^{\prime\prime}$ three mutually independent $\{0,1\}$-valued rvs
with success probability $p$. The two groups of rvs 
are assumed to be independent.

\section{First moments} \label{sec:FirstMom}

Fix $n=2,3, \ldots $ and $\myvec{\theta}$ in $\mathbb{R}_+ \times [0,1]$.
For both the unit circle and unit interval, the enforced independence 
assumptions readily imply
\begin{eqnarray}
\bE{ \chi_{n,1} (\theta) }
&=&
\bE{
\prod_{i=1,\ j \neq i}^n \left ( 1 - \chi_{ij} (\theta) \right )
}
\nonumber \\
&=&
\bE{ \left ( 1 - p a(X;r) \right )^{n-1} }
\nonumber \\
&=&
\int_0^1 \left ( 1 - p a(x;r) \right )^{n-1} dx
\label{eq:ExpressionsForFirstMoment}
\end{eqnarray}
where we have set
\begin{equation}
a(x;r) := \bP{ d( x ,  Y ) \leq r },
\quad
\begin{array}{c}
0 \leq x \leq 1 ,\\
r > 0 .
\end{array}
\label{eq:ExpressionsForA}
\end{equation}
Closed-form expressions for (\ref{eq:ExpressionsForA}) depend on the
geometric random graph being considered.

\subsection{The unit circle}

As there are no border effects, we get
\begin{equation}
a^{(C)}(x;r) = \ell(r), 
\quad 
\begin{array}{c}
0 \leq x \leq 1 ,\\
r > 0
\end{array}
\label{eq:AforCircle}
\end{equation}
and with the help of (\ref{eq:ExpressionsForFirstMoment})
this yields
\begin{equation}
\bE{ \chi^{(C)}_{n,1} (\myvec{\theta}) }
=  \left ( 1 - p \ell(r) \right)^{n-1},
\quad 
\begin{array}{c}
r > 0,\\
p \in [0,1].
\end{array}
\label{eq:FirstMomCircle}
\end{equation}

\subsection{The unit interval}

For $r \ge 1$, it is plain that
\begin{equation*}
a^{(L)} (x;r) = 1,
\quad 0 \leq x \leq 1 .
\end{equation*}
On the other hand, when $0 < r < 1$,
elementary calculations show that
\remove{
\begin{eqnarray*}
& & a^{(L)} (x;r) \\
& = & \left \{
\begin{array}{ll}
x + r & \mbox{if~ $0 \leq x \leq \min(r,1-r)$} \\
          &                \\
\ell(r) & \mbox{if~ $ \min(r,1-r) < x < \max(r,1-r) $} \\
          &                \\
1-x + r & \mbox{if~ $\max(r,1-r) \leq x < 1$.} 
\end{array}
\right .
\end{eqnarray*}}
\begin{eqnarray*}
& & a^{(L)} (x;r) \\
& = & \left \{
\begin{array}{ll}
x + r & \begin{array}{l}
\text{if } 0 < r \le 0.5, 0 \leq x \leq r \\
\text{or } 0.5 < r < 1, 0 \leq x \leq 1-r
\end{array} \\
&                \\
\ell(r) & \begin{array}{l}
\text{if } 0 < r \le 0.5, r \leq x \leq 1-r \\
\text{or } 0.5 < r < 1, 1-r \leq x \leq r
\end{array} \\
          &                \\
1-x + r & \begin{array}{l}
\text{if } 0 < r \le 0.5, 1-r \leq x \leq 1 \\
\text{or } 0.5 < r < 1, r \leq x \leq 1.
\end{array} \\ 
\end{array}
\right .
\end{eqnarray*}
Reporting this information into (\ref{eq:ExpressionsForFirstMoment}),
we obtain the following expressions in a straightforward manner:
\begin{itemize}
\item[(i)] For $0 < r \leq 0.5$ and $0 < p \leq 1$,
\begin{eqnarray}
\bE{ \chi^{(L)}_{n,1} (\myvec{\theta}) } 
&=& (1-2r)(1-2pr)^{n-1}
\label{eq:Expression(i)} \\
& & + ~ \frac{2}{np} \left ( (1-pr)^n-(1-2pr)^n \right ).
\nonumber 
\end{eqnarray}

\item[(ii)] For $0.5 < r < 1$ and $0 < p \leq 1$,
\begin{eqnarray}
\bE{ \chi^{(L)}_{n,1} (\myvec{\theta}) }
&=& (2r-1)(1-p)^{n-1}
\label{eq:Expression(ii)} \\
& & +~ \frac{2}{np} \left ( (1-pr)^n-(1-p)^n \right ).
\nonumber 
\end{eqnarray}

\item[(iii)] For $r \geq 1$ and $0 < p \leq 1$,
\begin{equation}
\bE{ \chi^{(L)}_{n,1} (\myvec{\theta}) } = (1-p)^{n-1}.
\label{eq:Expression(iii)}
\end{equation}

\item[(iv)] For $r > 0$ and $p=0$,
\begin{equation}
\bE{ \chi^{(L)}_{n,1} (\myvec{\theta}) } = 1.
\label{eq:Expression(iv)}
\end{equation}
\end{itemize}

The expressions (\ref{eq:Expression(i)}) and (\ref{eq:Expression(ii)})
can be combined into the single expression
\begin{eqnarray}
\bE{ \chi^{(L)}_{n,1} (\myvec{\theta}) }
&=& \left | 2r-1 \right | (1-p\ell(r))^{n-1}
\label{eq:Expression(i+ii)} \\
& & +~ \frac{2}{np}
\left ( (1-pr)^n-(1-p\ell(r))^n \right )
\nonumber
\end{eqnarray}
on the range $0 < r < 1$ and $0 < p \leq 1$.
Collecting (\ref{eq:Expression(iii)}), (\ref{eq:Expression(iv)})
and (\ref{eq:Expression(i+ii)})
we get the upper bound
\begin{equation}
\bE{ \chi^{(L)}_{n,1} (\myvec{\theta}) } 
\leq 
(1-p \ell(r))^{n-1} 
+ 
\frac{2}{np}\left(1-\frac{1}{2}p \ell(r)\right)^n
\label{eq:IntervalFirstMomUB}
\end{equation}
for any fixed $n=2,3,\dots,$
and $\myvec{\theta}$ in $\mathbb{R}_+ \times [0,1]$.

\section{Second moments}

Again fix $n=2,3, \ldots $ 
and $\myvec{\theta}$ in $\mathbb{R}_+ \times [0,1]$.
The same arguments apply for both the unit circle and unit interval:
For $x,y$ in $[0,1]$, write
\begin{eqnarray*}
& & b(x,y;\myvec{\theta})
\nonumber \\
&:=&
\bE{
\left ( 1 - B^\prime \1{ d(x,Z) \leq r } \right )
\left ( 1 - B^{\prime\prime} \1{ d(y,Z) \leq r } \right )
}
\nonumber \\
&=& 1 - p a(x;r) - pa(y;r) + p^2 u(x,y;r)
\label{eq:ExpressionsForB}
\end{eqnarray*}
with
\[
u(x,y;r)
:=
\bP{ d(x,Z) \leq r, d(y,Z) \leq r}.
\]

We then proceed with the decomposition
\begin{eqnarray}
& & \chi_{n,1}(\myvec{\theta}) \chi_{n,2} (\myvec{\theta}) 
\nonumber \\
&=& \prod_{j=2}^n \left ( 1 - \chi_{1j} (\myvec{\theta}) \right ).
\prod_{k=1, k \neq 2}^n \left ( 1 - \chi_{2k} (\myvec{\theta}) \right ) 
\nonumber \\
&=& \left ( 1 - \chi_{12} (\myvec{\theta}) \right )
\prod_{j=3}^n \left ( 1 - \chi_{1j} (\myvec{\theta}) \right )
\left ( 1 - \chi_{2j} (\myvec{\theta}) \right ) .
\nonumber
\end{eqnarray}
Under the enforced independence assumptions, 
an easy conditioning argument 
(with respect to the triple $X_1$, $X_2$ and $B_{12}$) 
based on this decomposition now gives
\begin{eqnarray}
& & \bE{ \chi_{n,1}(\myvec{\theta}) \chi_{n,2} (\myvec{\theta}) } 
\label{eq:SecondMom+DistMatters} \\
&=&
\bE{ \left ( 1 - B \1{ d(X,Y) \leq r } \right )
      b( X,Y; \myvec{\theta} )^{n-2} } .
\nonumber
\end{eqnarray}

As mentioned earlier we need only consider the unit circle as we do
from now on: From (\ref{eq:AforCircle}) it is plain that
\[
b^{(C)} (x,y;\myvec{\theta})
= 1 - 2p \ell (r) + p^2 u^{(C)}(x,y;r)
\]
for all $x,y$ in $[0,1]$, where we note that
\begin{eqnarray*}
u^{(C)}(x,y;r) & = & \bP{ \| x - Z \| \leq r, \| y - Z \| \leq r } \\
 & = & u^{(C)} ( 0, \| x-y \|; r ) 
\end{eqnarray*}
by translation invariance.
Thus, writing
\begin{equation}
\tilde b^{(C)} (z; \myvec{\theta} )
:= 1 - 2p \ell (r) + p^2 \tilde u^{(C)}(z;r),
\quad z \in [0,0.5]
\label{eq:bTildeuTilde}
\end{equation}
with
\[
\tilde u^{(C)} (z;r) 
:=  u^{(C)} ( 0, z; r ) ,
\]
we get
\[
b^{(C)} (x,y;\myvec{\theta})
= \tilde b^{(C)} (\| x-y \|;\myvec{\theta}),
\quad x,y \in [0,1] .
\]

Taking advantage of these facts we now find
\begin{eqnarray*}
& & \bE{ \chi_{n,1}^{(C)}(\myvec{\theta}) \chi_{n,2}^{(C)} (\myvec{\theta}) } 
\nonumber \\
&=&
\bE{ \left ( 1 - B \1{ \| X - Y \| \leq r } \right )
      \tilde b^{(C)} ( \| X - Y \| ; \myvec{\theta} )^{n-2} } 
\nonumber \\
&=&
\bE{ \left ( 1 - p \1{ \| X - Y \| \leq r } \right )
      \tilde b^{(C)} ( \| X - Y \| ; \myvec{\theta} )^{n-2} }
\nonumber \\
&=& 2 \int_0^{0.5}
\left ( 1 - p \1{ z \leq r } \right ) 
\tilde b^{(C)} (z ; \myvec{\theta} )^{n-2} dz 
\label{eq:chi1chi2IntegralA}
\end{eqnarray*}
by a straightforward evaluation of the double integral
\[
\int_0^1 dx \int_0^1 dy
\left ( 1 - p \1{ \| x - y \| \leq r } \right ) 
\tilde b^{(C)}(\| x-y \|;\myvec{\theta} )^{n-2} .
\]
Consequently,
\begin{equation}
\bE{ \chi_{n,1}^{(C)}(\myvec{\theta}) \chi_{n,2}^{(C)} (\myvec{\theta}) } 
\leq 
2 \int_0^{0.5}
\tilde b^{(C)} (z ; \myvec{\theta} )^{n-2} dz .
\label{eq:chi1chi2IntegralB}
\end{equation}

It is possible to compute the value of $\tilde u^{(C)} (z;r)$
%hence of $b^{(C)} (z; \myvec{\theta} )$ through \ref{eq:bTildeuTilde}), 
for various values for $z,r$:
For $0 < r < 0.5$, we find 
\begin{align*}
& \tilde u^{(C)} (z;r) \\
= & \left \{
\begin{array}{ll}
2r-z & \mbox{if~ $0 < r < 0.25, 0 \leq z \le 2r$} \\
          &                \\
0 & \mbox{if~ $0 < r < 0.25, 2r < z \le 0.5$} \\
					& 								\\
2r-z & \mbox{if~ $0.25 \le r < 0.5, 0 \leq z \le 1-2r$} \\
          &                \\
4r-1 & \mbox{if~ $0.25 \le r < 0.5, 1-2r < z \le 0.5$}.
\end{array}
\right .
\end{align*}
\remove{
\begin{align*}
& \tilde u^{(C)} (z;r) \\
= & \left \{
\begin{array}{ll}
2r-z & \mbox{if~ $0 \leq z \le \min(2r,1-2r)$} \\
          &                \\
(4r-1)^+ & \mbox{if~ $\min(2r,1-2r) < z < 0.5$}.
\end{array}
\right .
\end{align*}}
Details are outlined in Appendix \ref{sec:utilde}.

Obviously, if $r \geq 0.5$, then
$\tilde u^{(C)} (z;r)=1$ for every $z$ in $[0,0.5]$.
Thus, for $0 \le p \le 1$, through (\ref{eq:bTildeuTilde})
we obtain
\begin{align*}
& \tilde b^{(C)} (z; \myvec{\theta} ) \\
= & \left \{
\begin{array}{ll}
1-4pr + p^2 (2r-z) & \mbox{if~ $0 < r < 0.25, 0 \leq z \le 2r$} \\
          &                \\
1-4pr & \mbox{if~ $0 < r < 0.25, 2r < z \le 0.5$} \\
					& 								\\
1-4pr + p^2(2r-z) & \mbox{if~ $\begin{array}{l}
0.25 \le r < 0.5, \\
0 \leq z \le 1-2r
\end{array}$ } \\
          &                \\
1-4pr + p^2(4r-1) & \mbox{if~ $\begin{array}{l}
0.25 \le r < 0.5, \\
1-2r < z \le 0.5.
\end{array}$ }
\end{array}
\right .
\end{align*}
Using this fact in (\ref{eq:chi1chi2IntegralB}) 
and evaluating the integral, 
we obtain the following upper bounds,
see Appendix \ref{sec:chi1chi2UB} for details:

\begin{itemize}
\item[(i)] For $0 < r < 0.25$ and $0 < p \leq 1$,
\begin{align*}
& \bE{ \chi^{(C)}_{n,1}(\myvec{\theta}) \chi^{(C)}_{n,2} (\myvec{\theta}) } \\
\leq & (1-4r)(1-4pr)^{n-2} \\
 & \hspace{1em} + \frac{2(1-4pr)^{n-1}}{(n-1)p^2} \left ( \left ( 1 + \frac{2p^2r}{1-4pr} \right )^{n-1} - 1
\right ) .
\end{align*}

\item[(ii)] For $0.25 \leq r < 0.5$ and $0 < p \leq 1$,
\begin{align*}
& 
\bE{ \chi^{(C)}_{n,1}(\myvec{\theta}) \chi^{(C)}_{n,2} (\myvec{\theta}) } 
\\
\leq & (4r-1)(1-2pr)^{2(n-2)} \\
& \qquad + (2-4r)(1-4pr+2p^2r)^{n-2} .
\end{align*}

\item[(iii)] For $r \ge 0.5$ and $0 < p \leq 1$,
\[
\bE{ \chi^{(C)}_{n,1}(\myvec{\theta}) \chi^{(C)}_{n,2} (\myvec{\theta}) } 
= (1-p)^{2n-3}.
\]

\item[(iv)] For  $r > 0$ and $p=0$,
\[
\bE{ \chi^{(C)}_{n,1}(\myvec{\theta}) \chi^{(C)}_{n,2} (\myvec{\theta}) } = 1.
\]
\end{itemize}

Furthermore, combining these bounds with (\ref{eq:FirstMomCircle}), 
we obtain the following upper bound on
\[
R_n(\myvec{\theta})
:=\frac{ \bE{ \chi^{(C)}_{n,1}(\myvec{\theta}) 
                \chi^{(C)}_{n,2} (\myvec{\theta}) } }
         { \bE{ \chi^{(C)}_{n,1} (\myvec{\theta}) }^2 }
\]
in the various cases listed below.
\begin{itemize}
\item[(i)] For $0 < r < 0.25$ and $0 < p \leq 1$,
\begin{eqnarray}
R_n(\myvec{\theta}) & \leq & \frac{1-4r}{1-4pr} 
\label{eqn:bnd1} \\
& & +~ \frac{2}{(n-1)p^2} \left(
\left(1+\frac{2p^2r}{1-4pr}\right)^{n-1} - 1 
\right). 
\nonumber
\end{eqnarray}
\item[(ii)] For $0.25 \le r < 0.5$ and $0 < p \le 1$,
\begin{eqnarray}
R_n(\myvec{\theta}) & \leq
& \frac{4r-1}{(1-2pr)^2} 
\label{eqn:bnd2} \\
& & +~ 
(2-4r) \frac{(1-4pr+2p^2r)^{n-2}}{(1-2pr)^{2(n-1)}}. 
\nonumber
\end{eqnarray}
\item[(iii)] For $r \geq 0.5$ and $0 < p \leq 1$,
\begin{equation} 
R_n(\myvec{\theta}) = \frac{1}{1-p}.
\label{eqn:bnd3}
\end{equation}
\item[(iv)] For $r>0$ and $p=0$,
\begin{equation} R_n(\myvec{\theta}) = 1.
\label{eqn:bnd4}
\end{equation}
\end{itemize}

\section{Proof of the one laws}

As discussed in Section \ref{sec:FirstAndSecond}, the one law will be
established if we show that (\ref{eq:ToShow1}) holds.
Below we consider separately the unit circle and the unit interval.
In that discussion we repeatedly use the elementary bound
\begin{equation}
1 - x \leq e^{-x},
\quad x \geq 0.
\label{eq:B}
\end{equation}

\subsection{One law over the unit circle}

The one law over the unit circle reduces to showing 
the following convergence.

\begin{lemma}
{\sl
For any scaling 
$\myvec{\theta}: \mathbb{N}_0 \rightarrow \mathbb{R}_+ \times [0,1]$,
we have
\begin{equation*}
\lim_{n \rightarrow \infty} n\bE{ \chi^{(C)}_{n,1}(\myvec{\theta}_n) } = 0
\quad \text{if} \quad \liminfty{n} \alpha_n = +\infty
\end{equation*}
where the sequence $\alpha: \mathbb{N}_0 \rightarrow \mathbb{R}_+ $
is determined through (\ref{eq:DeviationUnitCircle}).
}
\label{lem:OneLawCircle}
\end{lemma}

\myproof
Fix $n=1,2, \ldots $ and in 
the expression (\ref{eq:FirstMomCircle}) substitute 
$(r,p)$ by $(r_n,p_n)$ according to the scaling
$\myvec{\theta}: \mathbb{N}_0 \rightarrow \mathbb{R}_+ \times [0,1]$.
We get
\begin{eqnarray*}
n\bE{ \chi^{(C)}_{n,1}(\myvec{\theta}_n) } 
&=&
n \left ( 1 - p_n \ell(r_n) \right)^{n-1} 
\nonumber \\
&=& n \left(1-\frac{\log n +\alpha_n}{n}\right)^{n-1}
\nonumber \\
&\leq&
n e^{-\frac{n-1}{n}(\log n + \alpha_n)}
\nonumber \\
&=& n^{\frac{1}{n}} e^{ - \frac{n-1}{n} \alpha_n}
\end{eqnarray*}
where the bound (\ref{eq:B}) was used.
Letting $n$ go to infinity we get the desired conclusion since
$\lim_{n \rightarrow \infty} \alpha_n = \infty$.
\myendpf

\subsection{One law over the unit interval}

A similar step is taken for the random intersection graph
over the unit interval.

\begin{lemma}
{\sl
For any scaling
$\myvec{\theta}: \mathbb{N}_0 \rightarrow \mathbb{R}_+ \times [0,1]$,
we have
\begin{equation*}
\lim_{n \rightarrow \infty} n\bE{ \chi^{(L)}_{n,1}(\myvec{\theta}_n) } = 0 
\quad \text{if} \quad \liminfty{n} \alpha^\prime_n = +\infty
\end{equation*}
where the sequence $\alpha^\prime: \mathbb{N}_0 \rightarrow \mathbb{R}_+ $
is determined through (\ref{eq:DeviationUnitInterval}).
}
\label{lem:OneLawInterval}
\end{lemma}

\myproof
Fix $n=1,2, \ldots $ and in 
the upper bound (\ref{eq:IntervalFirstMomUB}) substitute 
$(r,p)$ by $(r_n,p_n)$ according to the scaling
$\myvec{\theta}: \mathbb{N}_0 \rightarrow \mathbb{R}_+ \times [0,1]$.
We get
\begin{eqnarray}
n\bE{ \chi^{(L)}_{n,1}(\myvec{\theta}_n) } 
&\leq &
n \left ( 1 - p_n \ell(r_n) \right )^{n-1}
\nonumber \\
& & +~ 
\frac{2}{p_n}\left(1-\frac{1}{2}p_n \ell(r_n)\right)^n .
\nonumber 
\end{eqnarray}
%as we make use of (\ref{eq:FirstMomCircle}).

As in the proof of Lemma \ref{lem:OneLawCircle}, we can show that
\[
\lim_{n \rightarrow \infty} 
n \left ( 1 - p_n \ell(r_n) \right )^{n-1}
= 0
\]
under the condition
$\lim_{n \rightarrow \infty} \alpha^\prime_n = \infty$; details are 
left to the interested reader.
The desired conclusion will be established as soon
as we show that
\begin{equation}
\lim_{n \rightarrow \infty}
\frac{2}{p_n}\left(1-\frac{1}{2}p_n \ell(r_n)\right)^n
= 0
\label{eq:NeededLimit}
\end{equation}
under the same condition 
$\lim_{n \rightarrow \infty} \alpha^\prime_n = \infty$.

To do so, fix $n=1,2,\ldots $ sufficiently large so that
$\alpha^\prime_n \geq 0$ -- This is always possible under
the condition
$\lim_{n \rightarrow \infty} \alpha^\prime_n = \infty$.
On that range we note that
\begin{eqnarray*}
& & \frac{1}{p_n}\left(1-\frac{1}{2}p_n \ell(r_n)\right)^n 
\nonumber \\
&\leq& 
\frac{1}{p_n \ell(r_n)}
\cdot \left(1-\frac{1}{2}p_n \ell(r_n)\right)^n 
\nonumber \\
&\leq& 
\left ( 
\frac{2 (\log n - \log \log n) + \alpha^\prime_n}{n} 
\right )^{-1}
e^{-\log n + \log \log n - \frac{1}{2} \alpha^\prime_n }
\nonumber \\
&=& 
\frac{ \log n }
     { 2 (\log n - \log \log n) + \alpha^\prime_n }
e^{ - \frac{1}{2} \alpha^\prime_n }
\nonumber \\
&\leq& 
\frac{ \log n }
     { 2 (\log n - \log \log n) }
e^{ - \frac{1}{2} \alpha^\prime_n }
\end{eqnarray*}
upon using the fact $\ell(r_n) \leq 1$
and the bound (\ref{eq:B}).
Letting $n$ go to infinity we obtain (\ref{eq:NeededLimit}).
\myendpf

\section{Proof of the zero laws}

As observed earlier, when dealing with the zero law we need only concern
ourselves with the unit circle case. Throughout this section, we take 
$\myvec{\theta}: \mathbb{N}_0 \rightarrow \mathbb{R}_+ \times [0,1]$ 
and associate with it the sequence 
$\alpha: \mathbb{N}_0 \rightarrow \mathbb{R}_+ $
through (\ref{eq:DeviationUnitCircle}).
We now show (\ref{eq:ToShow2}) and (\ref{eq:ToShow3}) under the condition
$\lim_{n \rightarrow \infty} \alpha_n = - \infty$.
This will complete the proof of the zero laws. 

In the discussion we shall make use of the following elementary fact:
For any sequence $a: \mathbb{N}_0 \rightarrow \mathbb{R}_+$, 
the asymptotic equivalence
\begin{equation}
(1-a_n)^n \sim e^{-na_n}
\label{eq:C}
\end{equation}
holds provided 
$\lim_{n \to \infty} a_n = \lim_{n \to \infty} na^2_n = 0$. 

\subsection{Establishing (\ref{eq:ToShow2})}

The first step is contained in the following zero-law complement
of Lemma \ref{lem:OneLawCircle}.

\begin{lemma}
{\sl For any scaling
$\myvec{\theta}: \mathbb{N}_0 \rightarrow \mathbb{R}_+ \times [0,1]$,
we have
\begin{equation*}
\lim_{n \rightarrow \infty} n\bE{ \chi^{(C)}_{n,1}(\myvec{\theta}_n) } 
= \infty
\quad \text{if} \quad \liminfty{n} \alpha_n = -\infty
\end{equation*}
where the sequence $\alpha: \mathbb{N}_0 \rightarrow \mathbb{R}_+ $
is determined through (\ref{eq:DeviationUnitCircle}).
}
\label{lem:ZeroLawCircle}
\end{lemma}

\myproof
Fix $n=1,2, \ldots $ and in
the expression (\ref{eq:FirstMomCircle}) substitute
$(r,p)$ by $(r_n,p_n)$ according to the scaling
$\myvec{\theta}: \mathbb{N}_0 \rightarrow \mathbb{R}_+ \times [0,1]$.
As in the proof of Lemma \ref{lem:OneLawCircle} we start with the
expression
\begin{equation}
n\bE{ \chi^{(C)}_{n,1}(\myvec{\theta}_n) }
=
n \left ( 1 - p_n \ell(r_n) \right)^{n-1}.
\label{eq:ZeroLawCircleExpression}
\end{equation}

Under the condition $\lim_{n \rightarrow \infty } \alpha_n = -\infty$
we note that $\alpha_n = - |\alpha_n |$ for all $n$ sufficiently large,
say for all $n \geq n^\star$ for some finite integer $n^\star$.
Using (\ref{eq:DeviationUnitCircle}) we get 
$|\alpha_n| \leq \log n $ on that range by
the non-negativity condition $p_n \ell(r_n) \geq 0$. 
Therefore,
\begin{equation}
p_n \ell (r_n) \leq \frac{\log n}{n}
\quad \mbox{\rm and} \quad
n \left ( p_n \ell (r_n) \right )^2
\leq \frac{ (\log n )^2}{n}
\label{eq:pTimeslIneq}
\end{equation}
for all $n \geq n^\star$, and the equivalence (\ref{eq:C}) 
(with $a_n = p_n \ell (r_n)$) now yields
\begin{equation}
n \left ( 1 - p_n \ell(r_n) \right)^{n-1}
\sim n e^{- n p_n \ell (r_n) }
\label{eq:a}
\end{equation}
with
\begin{equation}
n e^{- n p_n \ell (r_n) }
=
n e^{- (\log n + \alpha _n ) } = e^{-\alpha_n} ,
\quad n=1,2, \ldots
\label{eq:b}
\end{equation}

Finally, letting $n$ go to infinity in (\ref{eq:ZeroLawCircleExpression})
and using (\ref{eq:a})-(\ref{eq:b}), we find
\[
\lim_{n \rightarrow \infty}
n \left ( 1 - p_n \ell(r_n) \right)^{n-1}
= \lim_{n \rightarrow \infty} e^{ - \alpha _n }
= \infty
\]
as desired under the condition 
$\lim_{n \rightarrow \infty } \alpha_n = -\infty$.
\myendpf

\subsection{Establishing (\ref{eq:ToShow3})}

The proof of the one-law will be 
completed if we establish the next result.

\begin{proposition}
{\sl For any scaling
$\myvec{\theta}: \mathbb{N}_0 \rightarrow \mathbb{R}_+ \times [0,1]$,
we have
\begin{equation*}
\limsup_{n \rightarrow \infty} R_n(\myvec{\theta}_n) \leq 1 
\quad \text{if} \quad \liminfty{n} \alpha_n = -\infty
\end{equation*}
where the sequence 
$\alpha: \mathbb{N}_0 \rightarrow \mathbb{R}_+ $
is determined through (\ref{eq:DeviationUnitCircle}).
}
\label{prop:CovarianceLimit}
\end{proposition} 

The proof of Proposition \ref{prop:CovarianceLimit}
is organized around the following simple observation:
Consider a sequence $a: \mathbb{N}_0 \rightarrow \mathbb{R}$
and let $N_1, \ldots , N_K$ constitute a partition of $\mathbb{N}_0$
into $K$ subsets, i.e., $N_k \cap N_\ell = \emptyset$ for distinct
$k,\ell=1, \ldots , K$, and $\cup_{k=1}^K N_k = \mathbb{N}_0$.
In principle, some of the subsets $N_1, \ldots , N_K$ may be either
empty or finite. For each $k=1, \ldots , K$ such that $N_k$ 
is {\em non}-empty, we set
\[
\alpha_k :=
\limsup_{ \substack{n \rightarrow \infty\\
                    n \in N_k } }
a_n 
= \inf_{ n \in N_k}
\left ( \sup_{m \in N_k:~ m \geq n } a_m 
\right )
\]
with the natural convention that $\alpha_k = -\infty $ when $N_k$ is finite.
In other words, $\alpha_k$ is the limsup for the subsequence
$\{ a_n , \ n \in N_k \}$.
It is a simple matter to check that
\begin{equation*}
\limsup_{n \rightarrow \infty} a_n
= {\max}^\star \left ( \alpha_k, \ k=1, \ldots , K \right ) 
\end{equation*}
with $\max^\star$ denoting the maximum operation over
all indices $k$ such that $N_k$ is non-empty.

\myproof
As we plan to make use of this fact with $K=4$, we write
\[
R_k := \limsup_{ \substack{n \rightarrow \infty\\
                    n \in N_k } }
R_n(\myvec{\theta}_n),
\quad k=1,\ldots , 4
\]
with
\begin{align*}
N_1 := & \{ n \in \mathbb{N}_0: \ 0 < r_n < 0.25, \ 0 < p_n \leq 1 \},
\\
N_2 := & \{ n \in \mathbb{N}_0: \ 0.25 \leq r_n < 0.5, \ 0 < p_n \leq 1 \},
\\
N_3 := & \{ n \in \mathbb{N}_0: \ 0.5 \leq r_n , \ 0 < p_n \leq 1 \}
\\
\intertext{and}
N_4 := & \{ n \in \mathbb{N}_0: \ r_n > 0, \ p_n = 0 \}.
\end{align*}
Therefore, we have
\begin{equation*} 
\limsup_{n \rightarrow \infty} 
R_n(\myvec{\theta}_n) = {\max}^\star (R_k, \ k=1, \ldots , 4 ) 
\label{eqn:limsup=max}
\end{equation*}
and the result will be established if we show that
\begin{equation*}
R_k \leq 1, \quad k=1, \ldots , 4 .
\label{eq:TOSHOW}
\end{equation*}
In view of the convention made earlier, we need only discuss
for each $k=1, \ldots , 4$,
the case when $N_k$ is countably infinite, as we do from now on.

The easy cases are handled first:
From (\ref{eqn:bnd4}) it is obvious that $R_4 = 1$. 
Next as observed before, 
(\ref{eq:pTimeslIneq}) holds for all $n$ sufficiently large 
under the condition $\lim_{n \to \infty} \alpha_n = -\infty$.
Since $\ell(r_n)=1$ for all $n$ in $N_3$, we conclude that
\[
\lim_{ \substack{n \rightarrow \infty\\
                    n \in N_3 } }
p_n = 0
\]
and the conclusion $R_3 =1 $ is now immediate from (\ref{eqn:bnd3}).
We complete the proof by invoking 
Lemmas \ref{lem:N1} and \ref{lem:N2} given next
which establish $R_1 \leq 1$ and $R_2 \leq 1$, respectively.
\myendpf

\begin{lemma}
{\sl Under the assumptions of Proposition \ref{prop:CovarianceLimit},
with $N_1$ countably infinite, we have $R_1 \leq 1$.
}
\label{lem:N1}
\end{lemma}

\myproof
Fix $n=2,3, \ldots $ and pick $(r,p)$ such that
$0 < r < 0.25$ and $0< p \leq 1$.
With $(\ref{eqn:bnd1})$ in mind, we note that
\begin{eqnarray}
& & \frac{2}{(n-1)p^2}
\left(\left(1+\frac{2p^2 r}{1-4pr}\right)^{n-1} - 1 \right) 
\nonumber \\
&=&
\frac{2}{(n-1)p^2}
\left( \sum_{k=0}^{n-1} {n-1 \choose k} \left(\frac{2p^2 r}{1-4pr}\right)^k
- 1 \right)
\nonumber \\
&=&
\frac{2}{(n-1)p^2}
\sum_{k=1}^{n-1} {n-1 \choose k} \left(\frac{2p^2 r}{1-4pr}\right)^k
\nonumber \\
&=&
\frac{4r}{1-4pr}
+
\frac{2}{(n-1)p^2}
\sum_{k=2}^{n-1} {n-1 \choose k} \left(\frac{2p^2 r}{1-4pr}\right)^k
\nonumber
\end{eqnarray}
and
we can rewrite the right handside of (\ref{eqn:bnd1}) as 
\begin{eqnarray}
& & 
\frac{1-4r}{1-4pr} 
+ \frac{2}{(n-1)p^2} 
\left(\left(1+\frac{2p^2 r}{1-4pr}\right)^{n-1} - 1 \right)
\nonumber \\
&=&
\frac{1}{1-4pr} 
+ \frac{2}{(n-1)p^2}
\sum_{k=2}^{n-1} {n-1 \choose k} \left(\frac{2p^2 r}{1-4pr}\right)^k
\nonumber \\
&\leq&
\frac{1}{1-4pr} 
+ \frac{2}{(n-1)}
\sum_{k=2}^{n-1} {n-1 \choose k} \left(\frac{2pr}{1-4pr}\right)^k
\nonumber
\end{eqnarray}
since $p^k \leq p^2$ for $k=2, \ldots , n-1$.
Therefore,
\begin{equation*}
R_n (\myvec{\theta})
\leq
\frac{1}{1-4pr}
+ \frac{2}{(n-1)}
\left ( 1 + \frac{ 2pr }{1 - 4 pr} \right )^{n-1} .
\end{equation*}

In this last bound, fix $n$ in $N_1$ and substitute $(r,p)$ 
by $(r_n, p_n)$ according to the scaling 
$\myvec{\theta}: \mathbb{N}_0 \rightarrow \mathbb{R}_+ \times [0,1]$.
Standard properties of the limsup operation yield
\begin{eqnarray*}
R_1
&\leq&
\limsup_{ \substack{n \rightarrow \infty\\
                    n \in N_1 } }
\left ( \frac{1}{1-4p_n r_n} \right )
\\
& &
+~ \limsup_{ \substack{n \rightarrow \infty\\
                    n \in N_1 } }
\left ( \frac{2}{(n-1)}
\left ( 1 + \frac{ 2p_n r_n}{1 - 4 p_n r_n} \right )^{n-1}
\right )
\nonumber
\end{eqnarray*}
and the desired result $R_1 \leq 1$ will follow if we show that
\begin{equation}
\limsup_{ \substack{n \rightarrow \infty\\
                    n \in N_1 } }
\left ( \frac{1}{1-4p_n r_n} \right )
= 1
\label{eq:N1+A}
\end{equation}
and
\begin{equation}
\limsup_{ \substack{n \rightarrow \infty\\
                    n \in N_1 } }
\left ( \frac{2}{(n-1)}
\left ( 1 + \frac{ 2p_n r_n}{1 - 4 p_n r_n} \right )^{n-1}
\right )
= 0 .
\label{eq:N1+B}
\end{equation}

To do so,
under the condition $\lim_{n \to \infty} \alpha_n = -\infty$ we
once again use the fact that (\ref{eq:pTimeslIneq}) holds
for large $n$ with $p_n \ell (r_n) = 2 p_nr_n$ for all $n$ in $N_1$. 
Thus,
\[
\lim_{ \substack{n \rightarrow \infty\\
                    n \in N_1 } }
p_n r_n = 0
\]
and the convergence (\ref{eq:N1+A}) follows.

Next, since $1+x \leq e^x$ for all $x$ in $\mathbb{R}$, we note
for all $n$ in $N_1$ that
\begin{eqnarray*}
& & \frac{2}{n-1} \left(1+\frac{2p_n r_n}{1-4p_n r_n} \right)^{n-1}
\nonumber \\
&=& \frac{2}{n-1} 
\left(
1+\frac{p_n \ell(r_n)}{1-2 p_n \ell(r_n)} 
\right)^{n-1}
\nonumber \\
&\leq& \frac{2}{n-1}
\left ( e^{ \frac{p_n \ell(r_n)}{ 1-2 p_n \ell(r_n)} }  \right )^{n-1}
= 2 e^{\beta_n}
\end{eqnarray*}
with
\[
\beta_n
:=  (n-1) \frac{p_n \ell(r_n)}{ 1-2 p_n \ell(r_n)} - \log (n-1).
\]
Thus, (\ref{eq:N1+B}) follows if we show that
\begin{equation}
\lim_{ \substack{n \rightarrow \infty\\
                    n \in N_1 } }
\beta_n = -\infty .
\label{eq:N1+C}
\end{equation}
From (\ref{eq:pTimeslIneq}) we get
\[
\beta_n
\leq \left ( \frac{n-1}{n} \right )
\frac{\log n + \alpha_n}{ 1-2 \frac{\log n}{n} }
- \log (n-1)
\]
for large $n$.
It is now a simple exercise to check that
\[
\lim_{n \rightarrow \infty}
\left ( \frac{n-1}{n} \right ) \frac{\log n }{ 1-2 \frac{\log n}{n} }
- \log (n-1)
= 0
\]
and the conclusion (\ref{eq:N1+C}) is obtained under the assumption
$\lim_{n \to \infty} \alpha_n = -\infty$.
\myendpf

\begin{lemma}
{\sl Under the assumptions of Proposition \ref{prop:CovarianceLimit},
with $N_2$ countably infinite, we have $R_2 \leq 1$.
}
\label{lem:N2}
\end{lemma}

\myproof
Fix $n=2,3, \ldots $ and pick $(r,p)$ such that
$0.25 < r \leq 0.5$ and $0< p \leq 1$.
From $(\ref{eqn:bnd2})$ we get
\begin{eqnarray}
R_n(\myvec{\theta}) 
&\leq& 
\frac{4r-1}{(1-2pr)^2} 
\nonumber \\
& & +~ 
\frac{2-4r}{ (1-2pr)^2} \frac{(1-4pr+2p^2 r)^{n-2}}{(1-2pr)^{2(n-2)}}
\nonumber \\
&=& \frac{4r}{ (1-2pr)^2} 
\left ( 1 - \frac{(1-4pr+2p^2 r)^{n-2}}{(1-2pr)^{2(n-2)}} \right )
\nonumber \\
& & +~ 
\frac{1}{ (1-2pr)^2} 
\left ( 2 \frac{(1-4pr+2p^2 r)^{n-2}}{(1-2pr)^{2(n-2)}}
-1 \right ) .
\nonumber 
\end{eqnarray}

Now fix $n$ in $N_2$ and substitute $(r,p)$ by $(r_n, p_n)$ according to 
the scaling 
$\myvec{\theta}: \mathbb{N}_0 \rightarrow \mathbb{R}_+ \times [0,1]$
in (\ref{eqn:bnd2}). 
As before, properties of the limsup operation yield
\begin{equation}
R_2 \leq R_{2c}
\left ( R_{2a} + R_{2b} \right )
\label{Inequality1}
\end{equation}
with
\[
R_{2a}
:=
\limsup_{ \substack{n \rightarrow \infty\\
                    n \in N_2 } }
\left ( 4r_n 
\left ( 
1 - \frac{(1-4p_n r_n+2p_n^2 r_n)^{n-2}}{(1-2p_n r_n)^{2(n-2)}} 
\right )
\right ),
\]
\[
R_{2b}
:=
\limsup_{ \substack{n \rightarrow \infty\\
                    n \in N_2 } }
\left ( 2 \frac{(1-4p_n r_n+2p_n^2 r_n)^{n-2}}{(1-2p_n r_n)^{2(n-2)}}
-1 \right )
\]
and
\[
R_{2c}
:=
\limsup_{ \substack{n \rightarrow \infty\\
                    n \in N_2 } }
\frac{1}{(1-2p_n r_n)^2} .
\]

As in the proof of Lemma \ref{lem:N1}, it is also the case here that
$R_{2c}$ exists as a limit and is given by
\[
R_{2c}
=
\lim_{ \substack{n \rightarrow \infty\\
                    n \in N_2 } }
\frac{1}{(1-2p_n r_n)^2}  = 1 ;
\]
details are omitted in the interest of brevity.

Next, we show that
\begin{equation}
\lim_{ \substack{n \rightarrow \infty\\
                    n \in N_2 } }
\frac{(1-4p_n r_n+2p_n^2 r_n)^{n-2}}{(1-2p_n r_n)^{2(n-2)}}
= 1.
\label{eq:Intermediary1}
\end{equation}
Once this is done, we see from their definitions that
$R_{2a} = 0$ and $R_{2b} = 1$, and the conclusion 
$R_2 \leq 1$ follows from (\ref{Inequality1}).

To establish (\ref{eq:Intermediary1}) we note that
\[
4p_n r_n - 2p_n^2 r_n
= p_n \ell(r_n) (2-p_n) \leq 2 p_n \ell(r_n) 
\]
and
\[
2p_n r_n = p_n \ell (r_n)
\]
for all $n$ in $N_2$. Now making use of (\ref{eq:pTimeslIneq})
we conclude that
\[
\lim_{ \substack{n \rightarrow \infty\\
                    n \in N_2 } }
\left ( 4p_n r_n - 2p_n^2 r_n \right )
=
\lim_{ \substack{n \rightarrow \infty\\
                    n \in N_2 } }
(n-2) \left ( 4p_n r_n - 2p_n^2 r_n \right )^2
= 0
\]
while
\[
\lim_{ \substack{n \rightarrow \infty\\
                    n \in N_2 } }
2 p_n r_n 
= \lim_{ \substack{n \rightarrow \infty\\
                    n \in N_2 } }
(n-2) \left ( 2p_n r_n \right )^2
= 0.
\]
By the equivalence (\ref{eq:C}) used
with $a_n = 4p_n r_n - 2p_n^2 r_n$
and $a_n = 2 p_n r_n$, respectively, we now conclude that
\begin{eqnarray}
& & \frac{(1-4p_n r_n+2p_n^2 r_n)^{n-2}}{(1-2p_n r_n)^{2(n-2)}}
\nonumber \\
&\sim&
\frac{
e^{- (n-2) \left ( 4p_n r_n - 2p_n^2 r_n \right ) }
}{
\left ( e^{ - (n-2) \left ( 2 p_n r_n \right ) } \right )^2
}
\nonumber \\
&=& e^{ 2 (n-2) \left ( p_n^2 r_n \right ) }
\label{eq:Intermediary2}
\end{eqnarray}
as $n$ goes to infinity in $N_2$.

Finally, for $n$ in $N_2$, because $\ell(r_n) = 2 r_n \geq 0.5$,
we get
\begin{eqnarray*}
2 (n-2) \left ( p_n^2 r_n \right )
&=& (n-2) \frac{ \left ( p_n \ell(r_n) \right )^2 }
               { \ell(r_n) }
\nonumber \\
&\leq& 2 (n-2) \cdot \left ( p_n \ell(r_n) \right )^2
\nonumber \\
&=& \frac{2(n-2)}{n} \cdot n \left ( p_n \ell(r_n) \right )^2
\end{eqnarray*}
so that 
\[
\lim_{ \substack{n \rightarrow \infty\\
                    n \in N_2 } }
2 (n-2) \left ( p_n^2 r_n \right ) = 0
\]
with the help of (\ref{eq:pTimeslIneq}).
The conclusion (\ref{eq:Intermediary1})
now follows from (\ref{eq:Intermediary2}), and the proof of
Lemma \ref{lem:N2} is complete.
\myendpf

\section{Simulation results}

\begin{figure*}[htp]
\centerline{
\subfigure[Fix $n=100$, $p=0.25$ and vary $r$]{\includegraphics[width=3.25in]{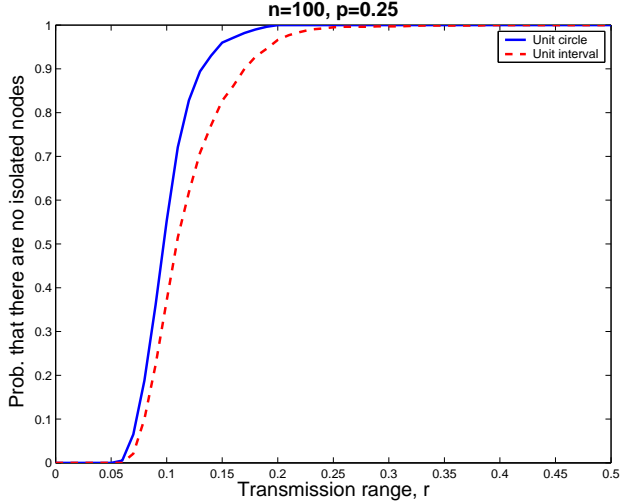}
\label{fig:sim-r}}
\hfil
\subfigure[Fix $n=100$, $r=0.1$ and vary $p$]{\includegraphics[width=3.25in]{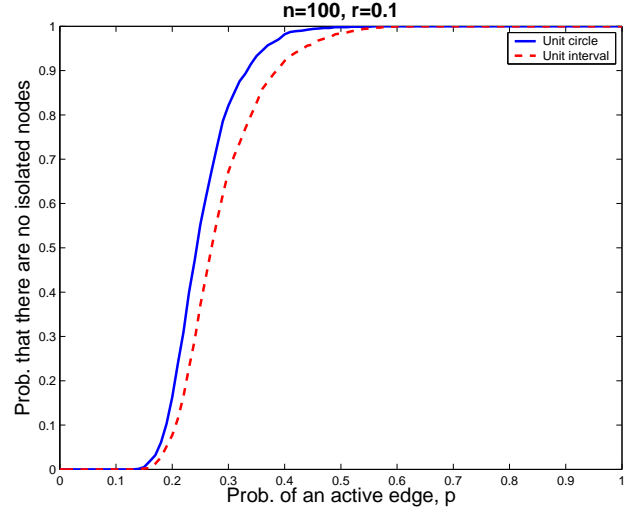}
\label{fig:sim-p}}}
\caption{Simulation results}
\label{fig:sim}
\end{figure*}

In this section, we present some plots from simulations in Matlab 
which confirm the results in Theorem \ref{thm:MainThmCir} 
and Theorem \ref{thm:MainThmInt}.
For given $n$, $p$ and $r$, we estimate
the probability that there are no isolated
nodes by averaging over $1,000$ instances 
of the random graphs $\mathbb{G}^{(C)}(n;\myvec{\myvec{\theta}})$ 
and $\mathbb{G}^{(L)}(n;\myvec{\myvec{\theta}})$.

In Figure \ref{fig:sim-r}, we have taken $n=100$ and $p=0.25$,
and examine the threshold behavior of the probability 
that there are no isolated nodes by varying $r$.
Theorem \ref{thm:MainThmCir} suggests that the critical range 
for the graph over the unit circle when $n=100$ and $p=0.25$ should be
$r^\star=0.09$. This is confirmed by the simulation results. 
In the case of the unit interval, we expect from 
Theorem \ref{thm:MainThmInt} that the critical range will be
between $r^\star=0.09$ and $r^{\star \star} =0.12$; this is in
agreement with the plot.

In Figure \ref{fig:sim-p}, we have taken $n=100$ and $r=0.1$,
and repeat the analysis by choosing various values for $p$. 
As expected from Theorem \ref{thm:MainThmCir},
the critical edge probability for the unit circle is found to occur
at $p^{\star} = 0.23$. It is also clear that for the unit interval, 
the critical edge probability is between
$p^{\star} = 0.23$ and $p^{\star \star} = 0.31$ as predicted by 
Theorem \ref{thm:MainThmInt}.

\section{Concluding remarks}
\label{ConcludingRemarks}

Theorem \ref{thm:MainThmInt} shows
a gap between the zero and one laws in the case of
the intersection graph on the unit interval:
The zero law expresses deviations
with respect to the scaling
$\myvec{\theta}^{\star}: \mathbb{N}_0 \rightarrow \mathbb{R}_+\times [0,1]$
determined through
\[
p^{\star}_n \ell(r^{\star}_n )
= \frac{ \log n }{n},
\quad n=1,2,\ldots
\]
as guessed.
On the other hand, 
the one law reflects sensitivity with respect to the \lq\lq {\em larger}"
scaling $\myvec{\theta}^{\star\star}: \mathbb{N}_0 \rightarrow 
\mathbb{R}_+\times [0,1]$
determined through
\[
p^{\star\star}_n \ell(r^{\star\star}_n )
= \frac{ 2(\log n - \log \log n )}{n},
\quad n=1,2,\ldots
\]
Inspection of the proof readily shows that the  
method of first moment is not powerful enough to close the gap --
To the best of our knowledge we are not aware of any other instance
in the literature where this occurs.
While we still believe that this gap can be bridged,
it is clear that a different method of analysis will be needed.

The analysis given here also suggests the form of the zero-one law to expect 
when the geometric component lives in higher dimensions. Specifically, 
consider the case where the nodes are located in a region 
$\mathbb{D} \subseteq \mathbb{R}^d$, 
without boundary, e.g., a torus or a spherical surface.
Then it is easy to compute the probability of an edge between two nodes as
\[
p \ell(r)= p \bP{ d(\myvec{x},\myvec{Y}) \leq r }
\]
where $\myvec{x}$ is an arbitrary point in $\mathbb{D}$, 
the rv $\myvec{Y}$ is uniformly distributed over $\mathbb{D}$
and $d(\cdot,\cdot)$ is the appropriate notion of distance. 
As before, if we define the sequence $\alpha: \mathbb{N}_0 \to \mathbb{R}$
through 
\[
p_n \ell(r_n) = \frac{\log n + \alpha_n}{n}, \quad n=1,2,\dots
\]
then the required dichotomy in the first moment 
(cf. Lemma \ref{lem:OneLawCircle} and Lemma \ref{lem:ZeroLawCircle})
cleary holds even in higher dimensions. As a result, we expect the 
critical scaling for the absence of isolated nodes to be given through
\[
p_n^\star \ell(r_n^\star) = \frac{\log n}{n}, \quad n=1,2, \dots.
\]

Finally, similar inferences can be made for modeling
wireless sensor networks which rely on the Eschenauer-Gligor scheme 
to securize their communication links: 
Power constraints restrict nodes to have a finite transmission range,
a physical communication constraint which is captured by the disk model,
the Eschenauer-Gligor scheme 
introduces a logical constraint which is well modeled by the random 
key graph \cite{YaganMakowskiISIT2008}. 
Combining these two constraints amounts to taking the intersection 
of a geometric random graph with a random key graph 
\cite{DiPietroManciniMeiPanconesiRadhakrishnan2004}
\cite{DiPietroManciniMeiPanconesiRadhakrishnan2006}.\footnote{The case
when the transmission range is inifinite is the so-called full visibility
case \cite{YaganMakowskiISIT2008}.}
However, unlike Erd\H{o}s-R\'enyi 
graphs, random key graphs exhibit dependencies between edges,
and this renders
the problem more complex. Nevertheless, we expect the determination
of critical scalings through the probability of an edge between two nodes 
to take place here as well; see (\ref{eq:CriticalRequirement}).
This time, in (\ref{eq:CriticalRequirement2})
the probability $p$ is replaced by the probability that two
nodes share a common key in the Eschenauer-Gligor scheme.

\appendix

\subsection{Calculation of $\tilde u^{(C)} (z;r)$} \label{sec:utilde}

Fix $0 < r < 0.5.$ With $X$ still denoting a rv uniformly distributed 
over $[0,1]$, we have
\begin{align}
& \tilde u^{(C)} (z;r) \nonumber \\
= & \bP{\|X\| \leq r, \|X-z\| \leq r} \nonumber \\
= & 1 - \bP{\|X\| > r} - \bP{\|X-z\| >r} \nonumber \\
 & \qquad \quad+ \bP{\|X\| > r, \|X-z\| >r}. \label{eq:utilde}
\end{align}
For the unit circle,
the probability that a uniformly distributed node falls
outside the range of a fixed node is independent of the node location,
hence
\[
\bP{\|X\| > r} = \bP{\|X-z\| >r} = 1-2r.
\]
Next, we consider
\begin{align*}
& \bP{\|X\| > r, \|X-z\| >r} \\
= & \bP{ \min(X,1-X) > r, \min(|X-z|, 1-|X-z|) > r } \\
= & \bP { E_1 \cap E_2 \cap (E_3 \cup E_4) } \\
= & \bP { E_1 \cap E_2 \cap E_3 } + \bP { E_1 \cap E_2 \cap E_4 } 
\end{align*}
where
\begin{align*}
E_1 & := \left[ r < X < 1-r \right], \\
E_2 & := \left[ z-(1-r) < X < z+(1-r) \right], \\
E_3 & := \left[ X > z+r \right] \\
\intertext{and}
E_4 & := \left[ X < z-r \right].
\end{align*}
It is clear that 
\begin{align*}
E_1 \cap E_2 \cap E_3 & = [ z+r < X < 1-r ] \\
\intertext{and}
E_1 \cap E_2 \cap E_4 & = [ r < X < z-r ]. 
\end{align*}

Consider the case $0< r <0.25$ and $0 \leq z \leq 2r$.
Then, the inequality
\begin{equation}
z \leq \min(2r, 1-2r),
\label{eq:zle2r}
\end{equation}
holds since $2r < 1-2r$ when $r <0.25$.
Therefore, 
\[
\bP{\|X\| > r, \|X-z\| >r} = 1-2r-z.
\]
Using this fact in (\ref{eq:utilde}), we obtain for
$0< r <0.25$ and $0 \leq z \leq 2r$ that
\[
\tilde u^{(C)} (z;r) = 2r - z.
\]
A similar calculation applies when
$0.25 \leq r < 0.5$ and $0 \leq z \leq 1-2r$
since (\ref{eq:zle2r}) holds in this case as well.

If $0 < r < 0.25$ and $2r < z \leq 0.5$, we obtain
\[
\bP{\|X\| > r, \|X-z\| >r} = (1-2r-z)+(z-2r) = 1-4r
\]
and this implies
\[
\tilde u^{(C)} (z;r) = 0 
\]
by substituting into (\ref{eq:utilde}). 

On the other hand, if $0.25 \leq r < 0.5$ and $1-2r < z \leq 0.5$
we get
\[
\tilde u^{(C)} (z;r) = 4r-1,
\]
since $\bP{\|X\| > r, \|X-z\| >r}=0$ in this case.

\subsection{Upper bound for 
$\bE{ \chi^{(C)}_{n,1}(\myvec{\theta}) \chi^{(C)}_{n,2} (\myvec{\theta}) }$}
\label{sec:chi1chi2UB}

The cases $(r\geq 0.5, 0 < p \leq 1)$ and $(p=0, r>0)$ are straightforward.
If $0 < r < 0.25$ and $0 < p \leq 1$, we use
(\ref{eq:chi1chi2IntegralB}) to obtain
\begin{align*}
& \bE{ \chi^{(C)}_{n,1}(\myvec{\theta}) \chi^{(C)}_{n,2} (\myvec{\theta}) } \\
\le & 2 \int_0^{2r} (1-4pr +p^2(2r-z))^{n-2} dz + 2 \int_{2r}^{0.5} (1-4pr)^{n-2} dz \\
= & \frac{2}{(n-1)p^2} \left( (1-4pr+2p^2r)^{n-1} - (1-4pr)^{n-1} \right) \\
 & \hspace{8em} + (1-4r)(1-4pr)^{n-2} \\
= & \frac{2(1-4pr)^{n-1}}{(n-1)p^2} \left ( \left ( 1 + \frac{2p^2r}{1-4pr} \right )^{n-1} - 1 \right ) \\
 & \hspace{8em} + (1-4r)(1-4pr)^{n-2}.
\end{align*}

For $0.25 \le r < 0.5$ and $0 < p \le 1$, we get
\begin{align*}
 & \bE{ \chi^{(C)}_{n,1}(\myvec{\theta}) \chi^{(C)}_{n,2} (\myvec{\theta}) } \\
\le & 2 \int_0^{1-2r} (1-4pr +p^2(2r-z))^{n-2} dz  \\
 & \hspace{4em} + 2 \int_{1-2r}^{0.5} (1-4pr+p^2(4r-1))^{n-2} dz \\
\le & (2-4r)(1-4pr+2p^2r)^{n-2} + (4r-1)(1-2pr)^{2(n-2)}. 
\end{align*}

\end{document}